\documentclass{article}

\usepackage{arxiv}

\usepackage[utf8]{inputenc} 
\usepackage[T1]{fontenc}    
\usepackage{hyperref}       
\usepackage{url}            
\usepackage{booktabs}       
\usepackage{amsfonts}       
\usepackage{nicefrac}       
\usepackage{microtype}      
\usepackage{lipsum}
\usepackage{graphicx}
\graphicspath{ {./images/} }

\usepackage{subcaption}
\usepackage{multirow} 
\usepackage{booktabs}
\usepackage{tabularx}
\title{DisTrack: a new Tool for Semi-automatic Misinformation Tracking in Online Social Networks}

\author{
 Guillermo Villar-Rodríguez \\
  Dpto. Sistemas Informáticos\\
  Universidad Politécnica de Madrid\\
  28031 Madrid, Spain\\
  \texttt{guillermo.villar@upm.es} \\
   \And
 Álvaro Huertas-García \\
  Dpto. Sistemas Informáticos\\
  Universidad Politécnica de Madrid\\
  28031 Madrid, Spain\\
  \texttt{alvaro.huertas@upm.es} \\
  \And
 Alejandro Martín \\
  Dpto. Sistemas Informáticos\\
  Universidad Politécnica de Madrid\\
  28031 Madrid, Spain\\
  \texttt{alejandro.martin@upm.es} \\
  \And
 Javier Huertas-Tato \\
  Dpto. Sistemas Informáticos\\
  Universidad Politécnica de Madrid\\
  28031 Madrid, Spain\\
  \texttt{javier.huertas.tato@upm.es} \\
  \And
   David Camacho \\
  Dpto. Sistemas Informáticos\\
  Universidad Politécnica de Madrid\\
  28031 Madrid, Spain\\
  \texttt{david.camacho@upm.es} \\
}

\begin{document}
\maketitle
\begin{abstract}
\textbf{Introduction:} This article introduces DisTrack, a methodology and a tool developed for tracking and analyzing misinformation within Online Social Networks (OSNs). DisTrack is designed to combat the spread of misinformation through a combination of Natural Language Processing (NLP) Social Network Analysis (SNA) and graph visualization. The primary goal is to detect misinformation, track its propagation, identify its sources, and assess the influence of various actors within the network.

\textbf{Methods:} DisTrack's architecture incorporates a variety of methodologies including keyword search, semantic similarity assessments, and graph generation techniques. These methods collectively facilitate the monitoring of misinformation, the categorization of content based on alignment with known false claims, and the visualization of dissemination cascades through detailed graphs. The tool is tailored to capture and analyze the dynamic nature of misinformation spread in digital environments.

\textbf{Results:} The effectiveness of DisTrack is demonstrated through three case studies focused on different themes: discredit/hate speech, anti-vaccine misinformation, and false narratives about the Russia-Ukraine conflict. These studies show DisTrack's capabilities in distinguishing posts that propagate falsehoods from those that counteract them, and tracing the evolution of misinformation from its inception.

\textbf{Conclusions:} The research confirms that DisTrack is a valuable tool in the field of misinformation analysis. It effectively distinguishes between different types of misinformation and traces their development over time. By providing a comprehensive approach to understanding and combating misinformation in digital spaces, DisTrack proves to be an essential asset for researchers and practitioners working to mitigate the impact of false information in online social environments.\end{abstract}


\section{Introduction}
\label{label:introduction}
At present, the impact of misinformation on our societies is beyond question. Governments, companies, researchers, among many others, are putting a great deal of effort into providing solutions and limiting the damage caused by this problem. The challenge, however, lies in the multifaceted nature of misinformation, making a comprehensive approach to combat it elusive. Addressing misinformation is contingent upon numerous factors, notably the underlying intent. Thus, when we use the word ``misinformation'', we are referring to information that is false by definition or unintentionally false information~\cite{salaverria2020desinformacion, said2021evolucion}, whereas ``disinformation'' is used when there is a deliberate intention to misinform.\cite{guess2020misinformation}, following the differences underlined by Karlova and Fisher in 2013~\cite{karlova2013social}. In addition to these terms, there is the concept of ``malinformation'', which groups every content created as a weapon~\cite{wardle2017information, ireton2018journalism}.

In order to understand the phenomenon of misinformation in the current era, and beyond intentionality, it is essential to include the means of transmission in the equation. As such, social media, particularly Online Social Networks (OSNs), has risen as society's primary channel for accessing information. Social media has transformed information dissemination from direct peer-to-peer exchanges to expansive many-to-many propagations, altering societal content consumption habits and thereby easing the spread of false information~\cite{posetti2018short}. Literature has demonstrated with specific cases how social media fosters falsehoods easily~\cite{kouzy2020coronavirus}. 

The combined effects of OSNs' dominance as information sources and the growing disenchantment with traditional news media amplify concerns about misinformation's influence on the public. According to the Reuters Institute Digital News Report, there's a notable decline in the practice of consulting a variety of news outlets and a general waning interest in news over the years. This trend includes the avoidance of topics like climate change and the Ukraine invasion by specific segments of the audience. This study underlines the contrast between worrying about misinformation and then depending more on social media to receive information~\cite{newman2023digital}.  

Despite the lack of expert consensus on the recent escalation of misinformation~\cite{posetti2018short}, academia has witnessed a significant surge in research articles tackling this issue, reflecting its growing impact~\cite{said2021evolucion, choras2021advanced}. In the years before the pandemic, these academic works had already increased exponentially~\cite{choras2021advanced}. Subsequent studies associate the surge in academic interest regarding misinformation, particularly from 2017, with events like the 2016 US elections~\cite{freelon2020disinformation}, suggesting these events catalyzed but did not solely trigger the focus~\cite{altaysurvey}. After these contributions to research, the emergence of more relevant international events and their originated misinformation highlights the importance of tackling this problem. 

In the fight against this phenomenon, fact-checkers represent the foremost defense. They are responsible for verifying thousands of pieces of information daily, cross-checking them and issuing statements on social media to counter the spread of false information. The International Fact-Checking Network (IFCN) built a ``Code of Principles'' to establish the criteria for the task of debunking misinformation. During the coronavirus crisis, the then Associate Director Cristina Tardáguila highlighted an unprecedented volume of misinformation, presenting a novel challenge for fact-checkers compared to prior years~\cite{brennen2020types}. Recent research on COVID-19 misinformation has shown the emergence of misinformation waves, with a plethora of hoaxes spreading rapidly, complicating efforts to address them all at once~\cite{martin2022facter}. Evidence like this demands new automated mechanisms not only at the level of verification but also in the following steps of mitigation on OSNs for a coordinated response against emerging falsehoods.

Mitigating the problem of misinformation therefore requires a complex approach where all factors of the problem are considered. In addition to the verification of the content itself, it is necessary to take into account how it is spread, the actors involved in its distribution and, in general, any element that participates in the cascade of misinformation propagation. In Fig.~\ref{fig:abstract_cascade} the described cascade is represented, where several actors interact over time about a given falsehood. We want to find and appropriately characterize this misinformation cascade, allowing for a better understanding of online discourse. As a result of considering all these factors, we draw the following research question for our work: \textit{“Is it possible to track conversations around specific hoaxes on Twitter (X)?”} which we sub-divide into four subquestions to answer it: 

\begin{figure}[!ht]
    \centering
    \includegraphics[width=0.75\textwidth,height=0.75\textheight,keepaspectratio]{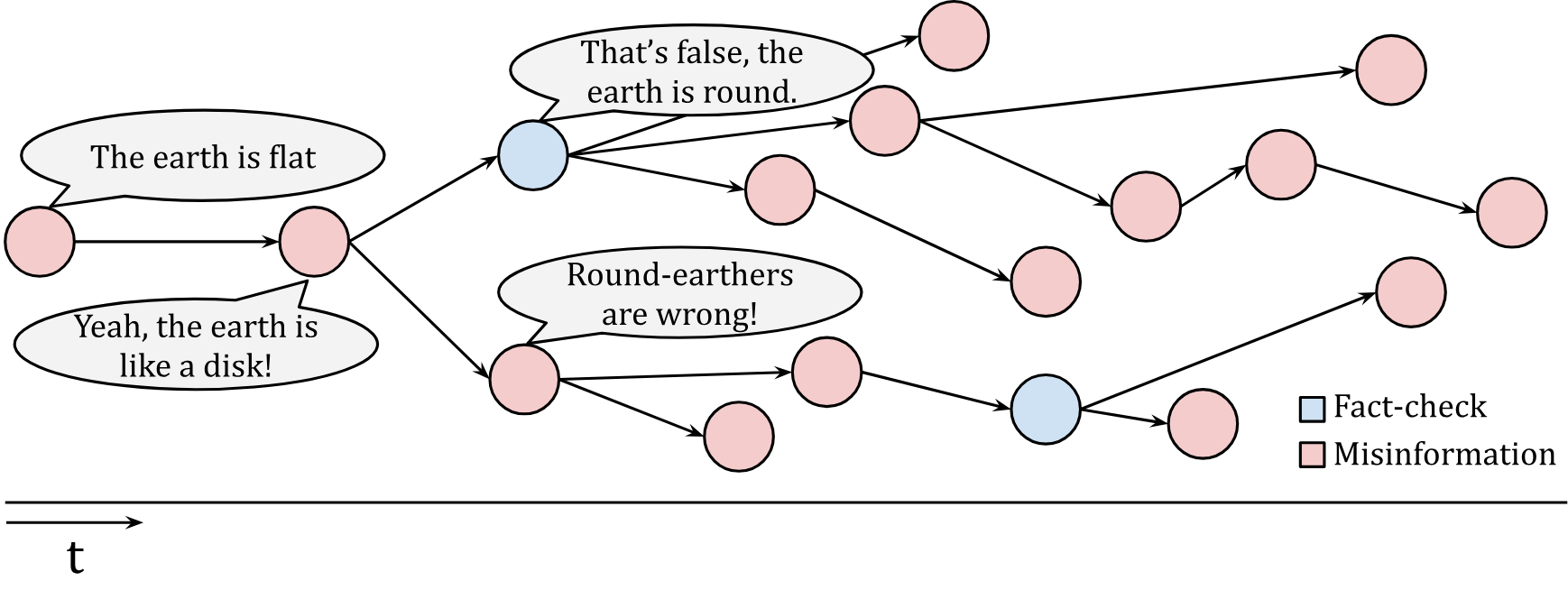}
    \captionsetup{justification=centering}
    \caption{Visualization of the misinformation cascade. The $t$ axis represents time from left to right, vertices are claims made by actors in any OSN, while edges represent a relation whether implicit (semantic similarity) or explicit (a repost of another piece of content).}
    \label{fig:abstract_cascade}
\end{figure}

\begin{enumerate}
    \item Can we extract the conversation about a hoax on Twitter (X)?
    \item Can we separate tweets related to the hoax in the extracted conversation from tweets not related to it?
    \item Can we distinguish between hoaxes that propagate a hoax from those that contradict it?
    \item Can we trace the movement of tweets related to a hoax with the users that spread them from beginning to end?
\end{enumerate}






We integrate cutting-edge Language Models and graph generation techniques to explore content propagation within Online Social Networks (OSNs) through the lens of Social Network Analysis (SNA). The social platform chosen for this objective is Twitter (X), given its dominant use for information: in proportion, it is more used than other social media for consuming mainstream news but also smaller or alternative news sources, and politicians and political activists, too~\cite{newman2023digital}. Furthermore, Twitter surpassed the other networks in this survey for the consumption of news about politics in the country and the Russia-Ukraine war, among others. This suggests the impact of false information and polarization could be disorders of these informative and political interests in this ecosystem and, thus, there would be a need to avoid them. 

This research enables not only the detection of misinformation but also the tracking of its journey from inception to conclusion within a Twitter (X) conversation, distinguishing between posts that propagate the falsehood and those that refute it. Our architecture is designed with two primary objectives: first, to \textbf{combat misinformation} by doing more than just verifying the veracity of a claim—we aim to identify every tweet that supports it; second, to \textbf{monitor and trace the sources of misinformation}. In the context of X, these sources of misinformation will be the tweets stating a false claim, not only the first tweets but all tweets that contribute to this, since all of them harvest this misinformation in the circles they belong to, whether the users inside them continue sharing it or not. This monitoring enables the analysis of the influence of various tweets and the identities of the individuals disseminating them.

Our methodology leverages a semi-automated fact-checking architecture~\cite{martin2022facter} that avoids the pitfalls of relying solely on a fully automated system. The traditional automated method, which categorizes information as true or false based on a pre-defined database, suffers from two major issues: it relies on a static collection of data that may not keep pace with emerging hoaxes, and it tends to categorize new, unverified information based on existing patterns, possibly mislabeling them. In contrast, our semi-automated strategy mitigates these issues by utilizing a dynamic database curated by fact-checkers that specifically targets false claims and can be regularly updated with new misinformation. Additionally, this method prioritizes the evaluation of claims in the database that align with the content of new information being assessed, ensuring a more accurate and timely verification process.

This integration of fact-checking organizations with AI techniques enables the automated identification of posts that are part of a hoax, as flagged in the fact-checkers' database. Distrack's semi-automated method advances Social Network Analysis by innovatively creating graphs that categorize the elements that create a cascade of information on social media according to their relationship with misinformation or with its debunking. This approach refines the analysis of viral content flows by filtering out irrelevant data, a significant departure from prior methodologies that did not distinguish between relevant and unrelated content. 

This article presents the following contributions:

\begin{itemize}
    \item \textbf{An integration of NLP and SNA to combat misinformation.} The study integrates cutting-edge language models and graph generation techniques to explore content propagation within online social networks (OSNs), specifically focusing on Twitter. 
    \item \textbf{A graph generation architecture for tracking misinformation.} The tool creates graphs that categorize the elements of information cascades on social media into three types: supporting a specific database claim, contradicting it, or being unrelated. This approach refines the analysis of viral content flows by filtering out irrelevant data and provides a comprehensive view of the spread of misinformation.
    \item \textbf{Case studies evaluation.} The paper presents three case studies that illustrate the application of DisTrack in tracking different types of false information related to topics such as COVID-19 vaccines, political issues, and more. These case studies demonstrate the tool's versatility and effectiveness in misinformation tracking across various subjects.
\end{itemize}

The product of this research is conceived as a cognitive computation system, according to its purpose of replicating human problem-solving~\cite{hasan2021review}. In this paper, it entails the human tasks of detecting false information, discovering all the posts related to it and analyzing the evolution of them, towards the final tracking of falsehoods. Through the computational methods presented, these manual processes are automated, assisted through AI.

More specifically, Natural Language Processing is cited as one of the areas in charge of cognitive systems and analyzing content in OSNs as one of the examples of computerized cognitive abilities. Likewise, the extraction of knowledge with Social Network Analysis is one of the techniques mentioned in Decision Support Systems (DSS), being these a combination of the outcomes of complex data analyses and machine learning~\cite{hasan2021review}.

This paper continues as follows: Background section reviews the key concepts and models used in this research; the next section describes the three modules in the methodology, the tweets extraction procedure, Natural Language Inference and graph generation; after this, the following section disentangles three examples in which this described methodology has been set into practice. The answers to the research questions are derived from these evaluated use cases, composing the Results and Discussion section that substantiates the final conclusions in the last part and the future lines of work.

\section{Background}
\label{label:background}

In this section, earlier contributions to the visualization of misinformation are explored. Discovering the veracity of a claim is crucial to visualizing the dissemination of misinformation because there are several lines of discourse surrounding any false-information claim, with fact-checkers on one side and misinforming actors on the other. It is expected that any information cascade about a hoax is surrounded by two different and opposing narratives.

Our research is motivated by earlier works performing semi-automated fact-checking using transformer-based language models, able to detect whether pieces of content (or claims) are either factually fake or have undetermined veracity. Following the previous rationale we motivate our techniques by exploring language models applied to fact-checking, as well as Social Network Analysis (or SNA) to accurately portray the dissemination of misinformation across online social networks (OSNs).


\subsection{Language Models}
\label{label:natural_language_processing}

The development of machine learning and deep learning models in the field of Natural Language Processing has made it possible to deal with complex tasks related to Natural Language Understanding (NLU). One of the most important steps was the emergence of language models with the introduction of the attention mechanism, leading to the development of Transformer models like BERT~\cite{devlin2018bert}, RoBERTa~\cite{liu2019roberta}, and XML~\cite{lample2019cross}. Unlike earlier embeddings (like word2vec~\cite{mikolov2013distributed} or Glove~\cite{pennington2014glove}), Transformer models generate vectorial representations using contextual information from neighboring words in the surrounding text, where each word is semantically informed by the sentence.

These advances opened the era of Language Models (LMs), architectures trained for tasks such as predicting the next word, but designed for multiple NLP problems. Among these, there are many scenarios where LMs can be deployed to fuel fact-checking processes, providing significant improvements over traditional machine learning methods~\cite{tretiakov2022detection}. For example, these models facilitate the automation of fact-checking by employing binary classification to discern false facts within the input. The state-of-the-art literature shows promising results in this line of work~\cite{tretiakov2022detection, jwa2019exbake}. Furthermore, this approach can be refined beyond simple binary outcomes by using varied labels to provide more nuanced distinctions between types of information~\cite{montoro2023fighting}.



\subsection{Automated fact-checking}
\label{label:llms_misinformation}

Misinformation is an ever-shifting issue. New pieces of misinformation may emerge as time passes, new narratives may become misinformation whereas older known hoaxes become real after an unexpected world event happens. Automated models trained without information retrieval techniques will inevitably become obsolete within the span of a few months. Allowing a model to retrieve information from trusted sources allows for proper decision making~\cite{montoro2023fighting}. In contrast to this, an alternative approach arises where the dataset is conceived as a knowledge base~\cite{vijjali2020two}. In this source, the data consists of textual statements containing verified falsehoods. Using these falsehoods, a model can compare an unverified claim against any verified falsehood and if there is any match, we can determine that the unverified claim also contains a falsehood. In this structure fact-checkers have a double role: they are responsible for the curation of the database, as well as the interpretation of the model output, giving complete control of the semi-automated model to fact-checkers and responsibility of its application~\cite{martin2022facter}.

The process has two steps: information retrieval (IR) and Natural Language Inference (NLI). For IR, one of the most commonly used methods is the calculation of the semantic distance between semantic embeddings~\cite{huertas2021civic, huertas2021countering}. This approach does not depend on a preliminary dataset of posts on the social network chosen to assess the veracity of an unseen post on that platform, allowing the classification of texts that belong to messaging environments in which full datasets are rarely obtained, such as WhatsApp~\cite{gaglani2020unsupervised}. The advances in this type of pipelines in the era of coronavirus encouraged research to focus on refining that knowledge base for that specific context~\cite{guo2020cord19sts}. Nowadays, tools in the fact-checking process based on the cosine similarity of texts have already been implemented in newsrooms and show their success over other methods~\cite{larraz2023semantic}.

The second step in this fact-checking process is determining the alignment between retrieved falsehoods and the original unverified content~\cite{vijjali2020two}. This alignment can be applied through the use of Natural Language Inference (NLI) as a subset of Natural Language. The task of NLI consists of checking if Sentence A (hypothesis) is inferred from Sentence B (premise)~\cite{maccartney2009natural}. In essence, this involves demonstrating that both sentences make the same assertion. The relationships obtained from NLI are the following: \textit{entailment}, when A and B refer to the same statement; \textit{neutral}, when A and B are not related to each other, and \textit{contradiction}, when A and B refer to the opposite statement~\cite{gururangan2018annotation}. 

The classification of statements in this task is performed by feeding the algorithms with datasets specifically designed for it. Stanford Natural Language Inference corpus (SNLI), with 570,000 pairs of sentences annotated with one of the three categories mentioned above~\cite{bowman2015large}, stands as the main reference for NLI, but further datasets solve the drawbacks of SNLI: for example, MultiNLI, with texts extracted from images cutlines, presents a more enriched text~\cite{williams2017broad}, or XNLI, with its cross-lingual approach, does not restrict NLI to one language~\cite{conneau2018xnli}. The use of NLI datasets enhanced with Transformers facilitates the comparison among languages~\cite{huertas2021sml}.

\subsection{Social Network Analysis}
\label{label:social_network_analysis}

Studying and mitigating the problem of misinformation involves detecting the misinformation itself, but also tackling the pathways by which it spreads. Thus, understanding how a piece of misinformation is disseminated on a social network is a vital tool. 

The flow of social media posts on Online Social Networks (OSNs) can be effectively represented as a graph. This graph-based structure organizes data into vertices linked by edges, providing a clear visualization of complex relationships~\cite{camacho2020four}. Within this framework, vertices represent either users or their posts, while edges illustrate the myriad interactions or relationships between them. These connections encompass the explicit social interactions derived from the network's metadata but, additionally, the more subtle, latent properties that link posts together.

Social Network Analysis (SNA) is the area in charge of studying these graphs from social platforms. The directions in this discipline comprise both the extraction of the common features of networks and the identification of aspects from the users from the graph~\cite{panizo2019describing}. Algorithms can be trained with this information, using, for example, the dynamics of likes~\cite{tacchini2017some}, to distinguish between types of posts. Misinformation arises as one of the emerging domains of application of SNA, together with politics and multimedia, being fields such as marketing, tourism, healthcare or cybersecurity more settled in this sort of studies~\cite{camacho2020four}.

Surveys in the field of misinformation have highlighted that the characteristics inside the post are one of the indicators to detect falsehoods, but the properties of the OSN itself play an important role. Sharma et al.~\cite{sharma2019combating} enumerated key elements such as the source/promoters, user responses and the information content parts. On the other hand, Parikh et al.~\cite{parikh2018media} specified non-text cues-based methods in the fight against false information, with user behaviour analysis as one of the subareas covered. In 2018, different studies demonstrated how false pieces of information were disseminated much further on Twitter than those that were true, by looking at properties such as the depth of the cascades generated by the post, the accounts contributing to spreading them and the duration of the propagation~\cite{vosoughi2018spread}. However, in the COVID-19 context, graph analyses revealed this expansion was only in terms of vastness: both false and legitimate information have the same influence, but actors spreading false information post more than those publishing the true one~\cite{saby2021twitter}.

However, regardless of the temporal context, repetitive patterns can be found. Before the coronavirus, SNA showed that individuals' decisions related to vaccines could be influenced inside circles debating about vaccination~\cite{bello2017detecting}, indicating that the connections among users on these platforms matter to the extent of dangerous implications if the communities approached are anti-vaccine. These contagions from one group to another bring the issue of virality, taken from the propagation of viruses. The creation of a cascade responds to one of the models of infection, the viral model where infected vertices by others can exchange the virus too, in contrast to the broadcast models in which contagions derive from a main vertex~\cite{goel2016structural}.

The role of the main spreaders in these cascades of misinformation among these circles has also been the focus of social-media-driven analyses. The change of information from one community to another through 'super-spreaders' and their characteristics were also disseminated with graphs~\cite{bodaghi2022theater}. Verified Twitter accounts (in the era before Elon Musk ownership and X) were shown to be 50 times more powerful in terms of propagating content about vaccines in comparison to non-verified profiles~\cite{carrasco2021participacion}.

To conclude, the use of NLP models without considering the dynamics of the social network allows us to visualize only part of the problem, leaving out key details~\cite{huertas2021civic, huertas2021countering, huertas2021sml}. Every instance of misinformation is surrounded by a community of users who interact with, share, comment on, support, or dispute it. Relying solely on content analysis limits our capabilities, overlooking the crucial task of unraveling the impact of false claims on Online Social Networks (OSNs), which extends beyond mere verification to include users' responses. The combination of NLP and SNA leads to a more realistic picture of the whole conversation of each falsehood, a technique yet to be sufficiently explored in the literature as we have seen in this section. 


\section{Tracking misinformation in OSNs} \label{label:monitoring_misinformation_in_graphs}
NLP and SNA can represent an alternative map of misinformation in OSNs through all the posts about a claim spreading a falsehood. In it, false information appears from viralization, but also from messages of different shapes from a wide to a short range of interactions and from a variety of users, not necessarily with the same impact in terms of their popularity in the social network. As an example, this approach allows to model the contagion from one vertex to the rest~\cite{goel2016structural} in two senses: on the one hand, NLP-driven research has demonstrated that there is not a unique message repeated in the propagation of misinformation, but many of them expressed differently; on the other hand, SNA-oriented studies show unconnected users outside the cascades also distribute falsehoods on social media.

Nevertheless, these studies contradicting the only focus on broadcast models~\cite{villar2022virality, noguera2023disinformation} in their combination of NLP and SNA are limited to the analysis of the data of the properties from the social network chosen in their final goal. Graph generation is ignored and the possibility of representing these ecosystems of falsehoods is missed. This results in an obstacle between the theory that confirms how misinformation is not just a cascade and the practice of representing what it is instead. This mentioned practice, in contrast to the previous approaches, would go further than the demonstration of a different model of diffusion of misinformation to reveal a more realistic flux of the posts causing it, examining its origin, evolution and end, if applicable.

\subsection{The DisTrack architecture}
The main goal of DisTrack is to create a complete representation of the propagation cascade of a piece of misinformation. To do so, it integrates different language models and SNA techniques that allow building a graphical representation of this cascade, providing information about the content and the actors in the social network that have played an essential role in the dissemination. DisTrack consists of three main sequential steps:

 \begin{enumerate}

 \item \textbf{Information retrieval from OSNs}: this module comprises the extraction of relevant keywords, the generation of queries through their possible combinations and the use of Twitter API to download all the tweets.
 \item \textbf{Semantic and Natural Language Inference}: this module refers to the conversion of tweets into Transformer-based embeddings that capture their meaning and context and the extraction of metrics based on their inference (if a tweet supports a false claim, contradicts it or is unrelated. For this second step, we make use of FacTeR-Check~\cite{martin2022facter}, which implements a semantic similarity filtering process followed by Natural Language Inference.
 
 \item \textbf{Graph generation}: this module ends with the hydration of the tweets downloaded to extract the insights to be used in the graph and the creation of it. Its vertices will be the tweets extracted and whose edges will correspond to the interactions among them. After this, the particularity of DisTrack is the use of the NLP-related and Twitter-related metrics to show the rest of the properties (position, size or colour).
 
\end{enumerate}

\begin{figure}[!ht]
    \centering    \includegraphics[width=0.85\textwidth,height=0.75\textheight,keepaspectratio]{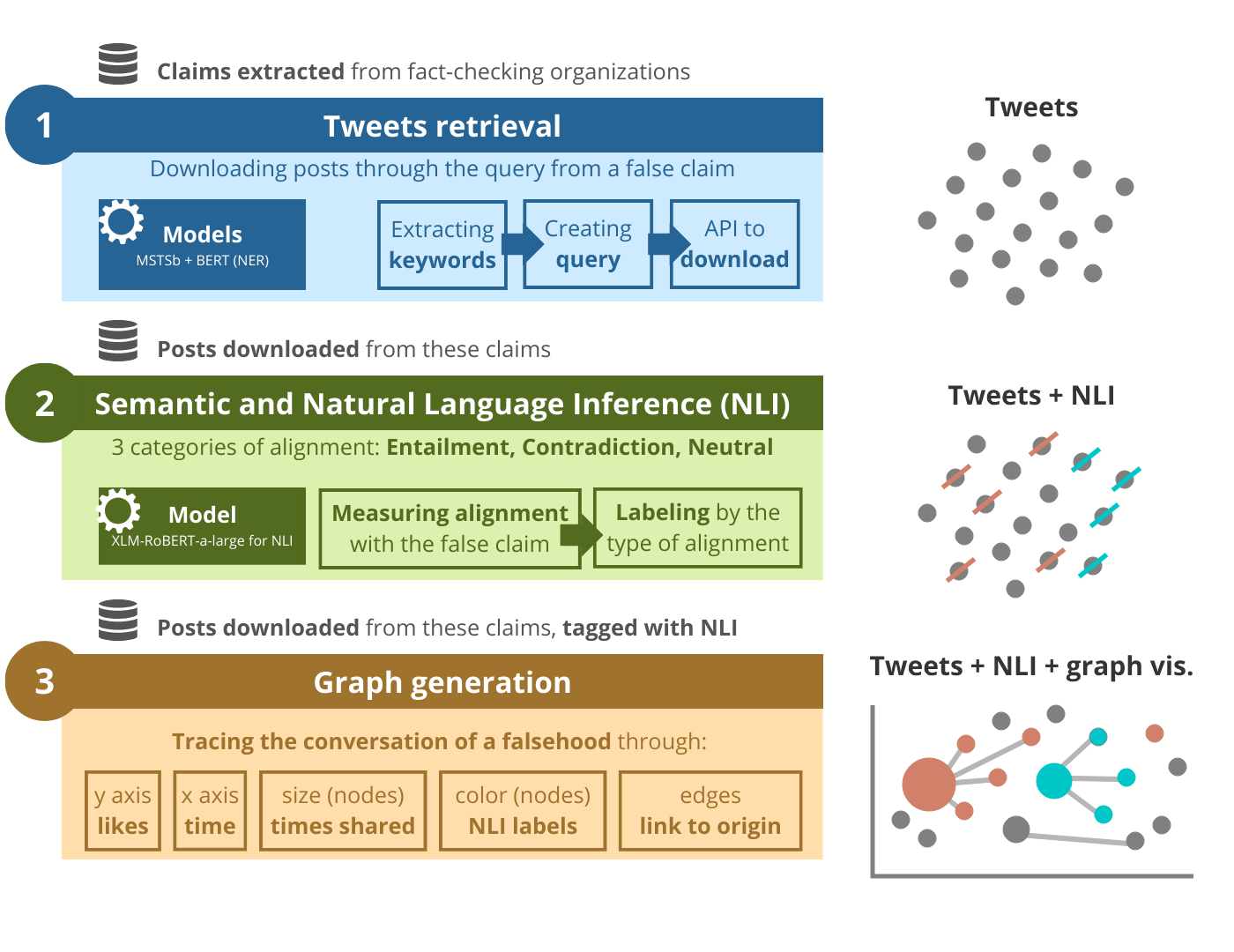}
    \captionsetup{justification=centering}
    \caption{DisTrack modules, top to bottom: 1) Information Retrieval, 2) Natural Language Inference, 3) OSN Tracking Visualization.}
    \label{fig:distrack-steps}
\end{figure}

DisTrack leverages the concept of tracking by modeling a graph based on a set of tweets downloaded from the OSN and labeled according to the alignment with a specific claim. This modeling process includes information extracted from the OSN such as following between users, retweets, or replies. As a result, the mechanisms of DisTrack can be distributed in three modules (see Fig.~\ref{fig:distrack-steps}).

The final output of these three modules is a visualization that acts as the operation center for the supervision of each piece of misinformation from the beginning to the end, assessing about the flux of a certain claim, the most influential spreaders and tweets with their connections or the periods in which this false information has had more impact.

This architecture allows us to characterize the discourse surrounding any piece of information. The visualization aims to highlight specific phenomena in social media discourse, as well as provide answers to questions for policymakers and content moderators. We provide some examples of these properties:
\begin{itemize}
    \item Proliferation (and decay) over time of falsehoods on social media. How does a piece of content propagate across time in an OSN?
    \item Impact of fact-checkers on the discourse surrounding a falsehood. Do they contribute to the proliferation or the decay of the surrounding discourse?
    \item Relationship between the influence of a social network actor and their impact on the discourse. How much do influential accounts dominate online discourse? Do smaller accounts have any impact on the evolution of the cascade?
    \item Detection of possible astroturfing campaigns. Has the falsehood spawned from several disconnected accounts? Is a coordinated attack on the social network being performed?
\end{itemize}

This division into modules gives them independence in performing a task. For example, the extraction of Twitter- and NLI-related insights without the need to follow the trajectory of false information has already been applied, without any further implementation, to discover the type of tweets, according to their interactions, that represent the true volume of misinformation beyond viral posts. These modules are explained in the subsections below.

To enable our research we begin exploring Twitter \footnote{now X, however we will call it Twitter for the sake of simplicity and understandability.}. This will be useful to evaluate our method on a heterogeneous set of use cases, but it is important to note that our methods can be applied to any social network that enables API search of textual content.

\subsection{Retrieving Twitter Content}
\label{label:module_for_queries_for_tweets_retrieval}
The process of extracting information from Twitter requires the execution of a series of searches for certain keywords. Since the Twitter API restricts the search to the exact keywords that are given as input, in order to retrieve a large representative sample of the relevant tweets and interactions related to one specific claim, it is necessary to build a set of multiple queries. By using different words and expressions in each of these queries it is possible to cover a large part of the tweets referring to the claim.


\subsubsection{Keyword search}
The query is the input to search and download tweets through Twitter API. These downloaded posts will contain the keywords inside that query. Through logical operators, queries can request the API tweets with every keyword inside it or for every tweet that at least has one of the keywords. However, this process is manual and the downloaded tweets are the result of the subjective decision of typing a query with a set of keywords instead of another. How these keywords are distributed with the logical operators to optimize the search is also an arbitrary decision. Furthermore, thinking of all these steps of converting a false claim into a query to download as many tweets as possible requires time, slowing down the computer-based fight against misinformation.

To solve these drawbacks of the manual creation of a query, an information retrieval module to generate queries and automate this process is needed. This research follows FacTer-ChecKey~\cite{martin2022facter}, a designed automated search to extract the most relevant keywords from a claim and concatenate them through the logical operator ``AND” to generate the final query. For example, the sentence ``Massive protest in France against the mandatory implementation of the COVID passport in public spaces” results in the concatenation of keywords ``(protest AND france AND passport AND covid AND public)”. This development takes inspiration from the model KeyBERT~\cite{grootendorst2020keybert} and it uses multilingual Transformer-based models to take the context from the meaning of a claim to extract the keywords from it, in addition to Name Entity Recognition (NER) to improve this extraction for those cases in Spanish, when needed.

This automated extraction faces the problem of the subjective decision of building the query manually, and they also deal with the subjective wording of the claim itself, from which the automated query is generated. In practice, including all the important keywords in these generated queries may not lead to the expected results because there can be tweets with some of the keywords from the claim but not necessarily all of them. Synonyms, abbreviations or word camouflage are examples of figures of speech contributing to a difficult keyword search.

We present a refined keyword extraction that tries to widen the search space and minimize the obstacles of this process, introducing three major add-ons: 
\begin{enumerate}
    \item The introduction of the logical operator \textit{OR} in the query, to allow multiple combinations of the keywords for the automated search
    \item The inclusion of a parameter that indicates the number of keywords that will be discarded from each combination of words, being the result of the query a concatenation of all the possible mixtures excluding two different keywords in each of them. In the example previously seen, the query would be optimized in this way: ``((protest AND france AND passport AND covid) OR (protest AND france AND passport AND public) OR (protest AND france AND covid AND public) OR (protest AND passport AND covid AND public) OR (france AND passport AND covid AND public))”. 
    \item The conversion of numerical numbers into all the possible forms, also textual, to avoid ignoring pieces of misinformation that include figures cited in a different way (e.g., ``10000 OR 10,000 OR 10.000 OR `10 thousand’ OR `ten  thousand’”)
\end{enumerate}

In this case, \verb|MSTSb-paraphrase-multilingual-mpnet-base-v2| model is used in the query generator in parallel with FacTer-Check, to encode the meaning together with the context of the words, and includes the model \verb|bert-spanish-cased-finetuned-ner| for Name Entity Recognition (NER). For this implementation, adverbs, conjunctions, adpositions and other stop words have been removed for the final composition with the multilingual Flair tagger and Spacy models.

Although the default settings could capture how a certain hoax has been diffused massively in a short time-lapse. For the sake of a better understanding of social media discourse, we are interested in extracting the maximum possible number of texts, actors and connections involved in the early steps and also the evolution of every falsehood, given the dimensions of misinformation on social media nowadays.

\subsubsection{Technical details}
The query generated through the claim of each piece of misinformation constitutes the input of Twitter API, whose access is offered through a developer account on Twitter. 

The reduced limitations granted by the academic API access soften two restrictions: the maximum number of tweets extracted and the time of the start of the final download. This last aspect can be modified to obtain information from a certain timestamp. This becomes crucial in the field of misinformation because the search must be constrained after the birth of the topic that is referenced in each hoax (for example, claims about COVID will not be found prior to the emergence of coronavirus).

Overall, the created query, the credential to validate the permissions (the token found on the developer account) and the selected time-lapse and maximum number of posts as parameters are included for the automated request for the needed tweets about misinformation. 
The data of each downloaded tweet consists of a JSON with Twitter-based information structured as metadata in different fields. 

However, since this cannot be exploited enough, NLP-based features in the following step will contribute to filtering non-related hoaxes and developing the final tool. These technical details describe a fixed frame of the use of the API at the moment of the experiment, but the current and future restrictions of this developers' system do not distort the steps of this research if the API is not available. Twitter versions of the API are a hands-on automated solution based on X advanced search, available through the search bar on the interface. What the API returns in terms of content is the same as the results that the interface gives to the user, also with the same query as input. For this reason, parameters such as the chosen time-lapse can be selected on X advanced search too.

The difference involves the codification of the output: whereas the API offers the massive download of tweets in JSON format, the advanced search gives the visual integration of those posts in the interface, which would need further processing to transform them as raw data and to structure their metadata.

\subsection{Automated Verification}
\label{label:module_for_semantic_and_natural_language_inference}

Once we have retrieved a sample of information from the OSN related to the input claim, each tweet is labeled according to the alignment with the original claim. We define alignment as the result of performing Natural Language Inference over a pair of content (the retrieved tweet from the last step) and falsehood (the piece of misinformation that is being tracked). After evaluating each retrieved Tweet we assign their labels.

\subsubsection{Natural Language Inference}

Natural language inference (NLI) is crucial to distinguish if a tweet is aligned or contradicts a false claim. Again, Transformer-based architectures are applied to measure this alignment regardless of how different each tweet is formulated in comparison to the false claim selected.
The NLI task consists of discovering if a hypothesis $h$ can be inferred from the premise $p$ in a pair of texts $(p, h)$. In the misinformation and fact-checking domains, $p$ will be each tweet from the conversation downloaded from Twitter and $h$ will be the piece of misinformation debunked by fact-checkers. Thus: $h$ is $h_f$ when stating that falsehood, and $h$ is $h_u$ when that factuality is undetermined.

The classification of posts according to their content through NLI is: 
\begin{itemize}
\item \textbf{Entailment} (when the falsehood is enunciated): a post that entails a piece of misinformation is a post that supports it and spreads it.
\item \textbf{Contradiction} (when the negation of the falsehood is enunciated): a post that contradicts a piece of misinformation is a post that denies it and may act as a protective shield in the circles where it arrives.
\item \textbf{Neutral} (when the falsehood or its contradiction is not enunciated): a post that is neutral bears no relevance to the discourse whatsoever a secondary effect of widening the spread of the keyword search.
\end{itemize}




\subsubsection{Technical details}
The semantic search uses the methodology proposed in FacTer-Check step by step without ad-dons or modifications. For the NLI task, a fine-tuned \verb|XLM-RoBERTa-large|~\cite{conneau2019unsupervised} was used for this module. For the NLI task, the Machine Translated MultiNLI (MNLI-MT)~\cite{williams2017broad} and XNLI~\cite{conneau2018xnli} datasets were used. Additional datasets such as ANLI~\cite{nie2019adversarial}, SNLI~\cite{bowman2015large} and FEVER~\cite{thorne2018fever} for English have been also included, using two training processes (inspired by FacTeR-Check): one only with MNLI (for the cross-lingual texts) and one with all the mentioned datasets. The hyperparameters chosen are: $1024$ as batch size, $2e$-$5$ as the learning rate, with Adam~\cite{kinga2015method} as the optimizer. Same for warmup and linear decay. The validation data in XNLI determines the optimal selected network after a manual hyper-parameter finetuning\footnote{Model available at \href{https://huggingface.co/AIDA-UPM/xlm-roberta-large-snli_mnli_xnli_fever_r1_r2_r3}{AIDA-UPM/xlm-roberta-large-snli\_mnli\_xnli\_fever\_r1\_r2\_r3} }.

\subsection{Graph visualization}
\label{label:module_for_graph_generation}

Our main contribution lies in the graph visualization module where we conceive how to translate an interconnected graph of tweets into a useful visualization of online discourse around a topic. Our DisTrack architecture makes use of all the properties stored in each tweet to make a readable composition, clearly showing how any piece of misinformation has evolved.

\subsubsection{Cascade graph building}
The information stored for each tweet includes the preceding author and the preceding tweet from which it derives. This information is enough to generate a directed graph $G = (V, E)$, with $V$ the vertex set containing tweets published by an author. In this graph, the $E$ edges represent a connection between vertices due to being either a reply, quote or retweet from another tweet, which will be the parent vertex. 

Each vertex has additional information contained by the contextual metadata of the author, likes among other details. Conveying this metadata to make it understandable is a non-trivial challenge that we address in the following subsections. In particular:
\begin{itemize}
    \item \textit{Time} is contained within the metadata, but the information is so unevenly spaced over time that the interpretation of the cascade is compromised.
    \item The cascade of misinformation is perpetrated mainly by actors that influence online discourse. Representing their \textit{influence} accurately in the visualization aside from the vertices is crucial to understanding the actual impact of actors.
    \item The information cascade of misinformation is twofold, containing usually two opposing narratives, the misinformation itself and the fact-checkers combating it. Greater insight can be achieved by using the results of NLI and integrating them into the visualization to evaluate the \textit{veracity} of the evaluated claims.
\end{itemize}

\subsubsection{Non-linear time representation}
The X-axis is defined by time. Each vertex will be placed according to the moment it has been published. However, time itself is not represented linearly in this axis. Falsehoods act in waves, the flow of information may be quick and sudden during a short time-lapse generating concentrated content, preventing the user from reading the evolution of that piece of false information properly when it explodes in terms of impact. Furthermore, if there is much distance in time from one vertex to the next one, again the space in the graph between these two points will create empty areas in the plot whereas the areas filled would be crowded with vertices (during the propagation of a hoax, there are moments of inactivity or a lower number of publications about it). Our approach just sorts posts chronologically, regardless of the time passed between one vertex and the next one, and sets different time stamps on the x-axis that improve the understanding of the propagation cascade.

\subsubsection{Author influence representation}
Regarding the y-axis, it measures the degree of influence of the author of a vertex/tweet. The number of followers allows us to distinguish between the influence of the different actors responsible for the spread of falsehood or their contradiction. 

However, the achieved impressions by the tweet are also affected by the number of likes, in addition to retweets and quotes, which are already represented through the children vertices. For this reason, likes will affect the size of each vertex proportionately in the final visualization. Obtaining this would result in a visualization that shows the evolution of a conversation of the tweets extracted from a query but also the explanation of it through the impact that users, retweets and likes generate.

\subsubsection{Veracity representation}
Finally, the outputs of NLI will serve as the colours to differentiate posts containing misinformation from those that contradict or are unrelated. This is the advance that allows us to check the beginning, transformation and current status of a hoax rather than just all the tweets downloaded through the main keywords appearing in a specific hoax, and that would culminate in the goal of tracing the conversations about misinforming posts, if the hypothesis is confirmed.

\section{Case studies}
\label{case_studies}

Three use cases illustrate the application of DisTrack. They contain the beginning and evolution of a piece of misinformation on Twitter, all of them with a different topic to show the versatility of this tool. Firstly, an exploratory data analysis of the downloaded tweets is made to disentangle the types of tweets and of their authors according to their NLI- and Twitter-based metrics. After this, as the proof of the variety of tweets contributing to the expansion of content related to misinformation, the representation of the final graphs is made thanks to the final module of graph generation from DisTrack.


These three cases represent three different topics:
\begin{itemize}
    \item \textbf{Case 1: ``The 80 percent of Muslims living in Europe live from social welfare
and they refuse to work''}. The first case is linked to the disbelief in institutions and hate against Muslims. It constitutes 32 original tweets and a positive balance of tweets involving \textit{entailment} in contrast to \textit{contradiction} (i.e., denying the hoax). The weight of \textit{entailment} increases much more when every post is shared. Thus, this case contains a total of 84 representative posts, with all the retweets included.

    \item \textbf{Case 2: ``RNA vaccines against coronavirus includes graphene oxid''}. It focuses on COVID-19-related antivaccine statements. It also includes 32 original tweets and a balanced number between \textit{entailment} and \textit{contradiction}. This second case has a total of 128 representative posts, including retweets.

    \item \textbf{Case 3: ``Zelensky sold 17 million hectares of land to Monsanto, Dupont and Cargill''}. It is an example of misinformation around the Russia-Ukraine war. It departs from 26 original tweets and most of the total tweets represent \textit{entailment} with the hoax (80\%). This third case involves a total of 916 tweets.
\end{itemize}

 \begin{figure}[!ht]
    \centering
    \includegraphics[width=0.85\textwidth,height=0.95\textheight,keepaspectratio]{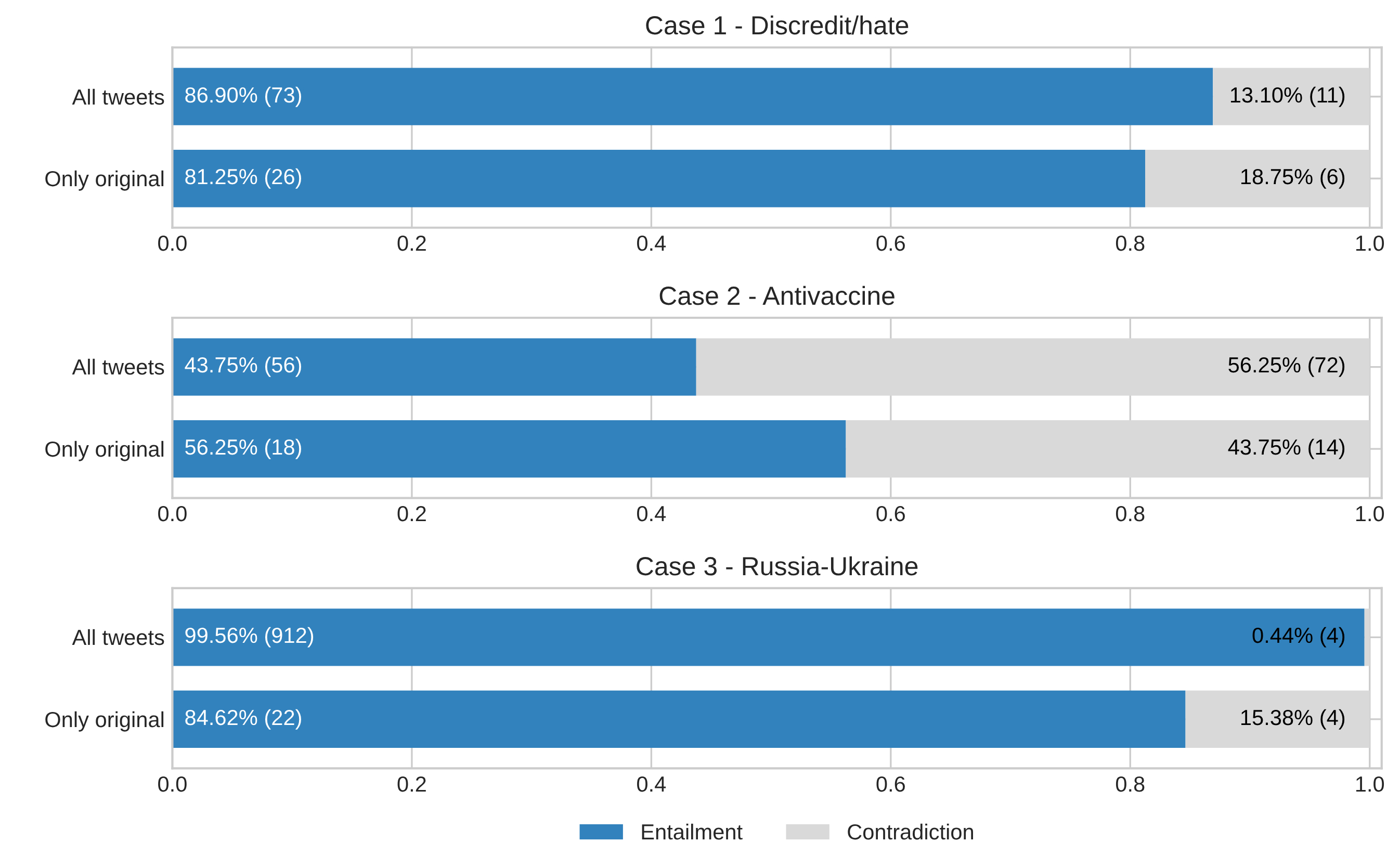}
    \captionsetup{justification=centering}
    \caption{Distribution of the groups of posts according to their type of NLI-based alignment (\textit{entailment} or \textit{contradiction}), in each of the three cases.}
    \label{fig:entailment_proportion}
\end{figure}

The specific distribution of the number of tweets supporting or denying each false claim is represented in Figure~\ref{fig:entailment_proportion}. Cases 1 and 3 are examples of the general trend that can be observed in the dissemination of false information. The presence of posts that support misinformation is much higher than those that contradict the news, mostly due to the activity of fact-checker accounts. In contrast, during the pandemic, we could see how a large part of the social media community actively participated in countering false information, as in the example of case 2.

 \begin{figure}[!ht]
    \centering
    \includegraphics[width=0.85\textwidth,height=0.95\textheight,keepaspectratio]{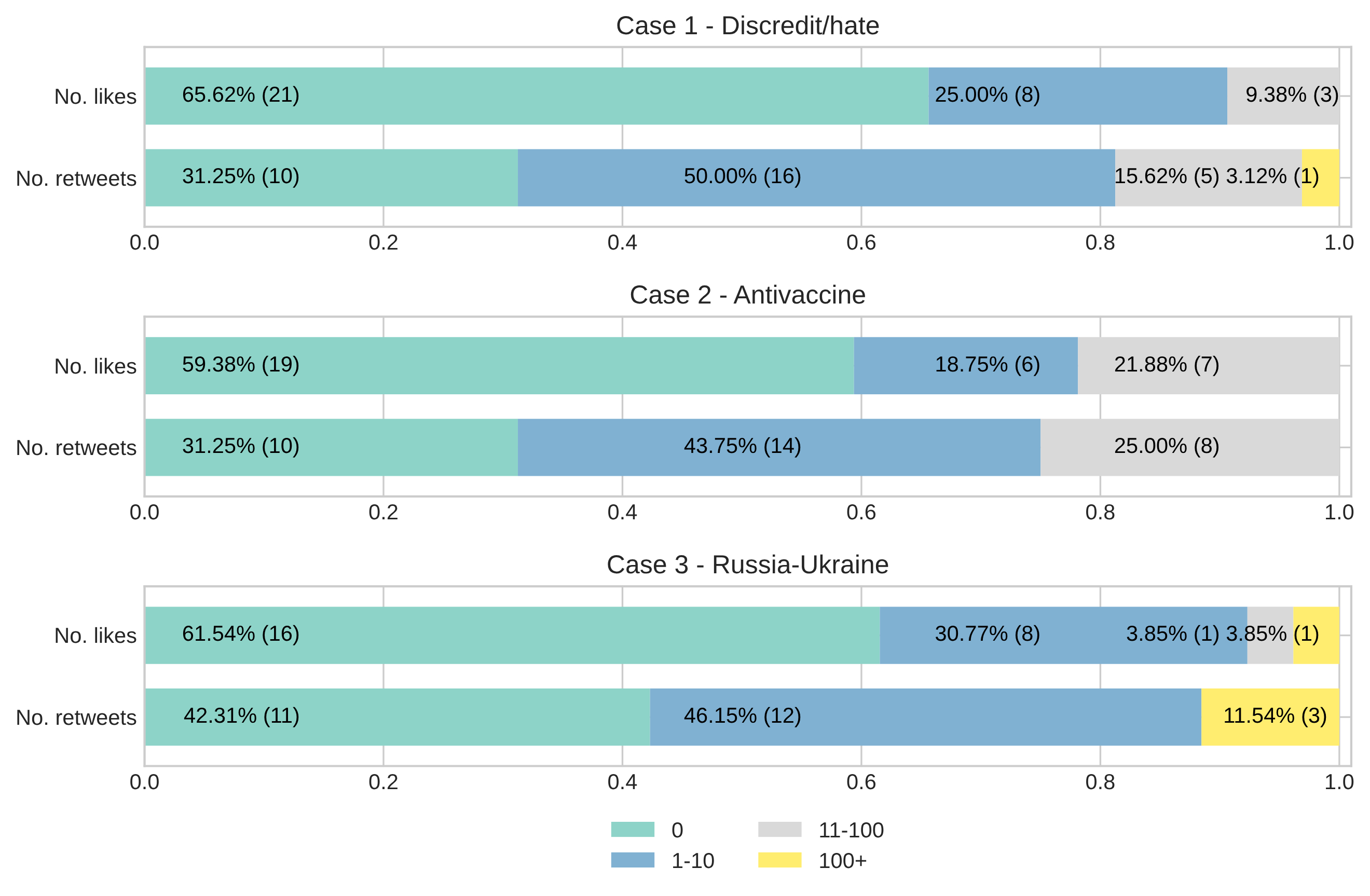}
    \captionsetup{justification=centering}
    \caption{Distribution of the groups of posts according to their number of retweets and number of likes, in each of the three cases, with only original tweets into account.}
    \label{fig:likes_retweets}
\end{figure}

In order to evaluate the impact of a post on a social network, the number of retweets or likes are two of the most commonly used indicators. As can be seen in Fig.~\ref{fig:likes_retweets}, in terms of retweets, tweets from one to ten reposts are the main group, followed by those with zero retweets. Regarding likes, most of the posts in the flux of misinformation do not contain any interaction, in contrast with the lower percentages of tweets between one and ten likes. Tweets between one and ten likes are a small proportion except in the case of the hoax related to the vaccines, and only the propagation of the misinformation cited about the war between Russia and Ukraine has a tweet with more than 100 likes.

 \begin{figure}[!ht]
    \centering
    \includegraphics[width=0.85\textwidth,height=0.95\textheight,keepaspectratio]{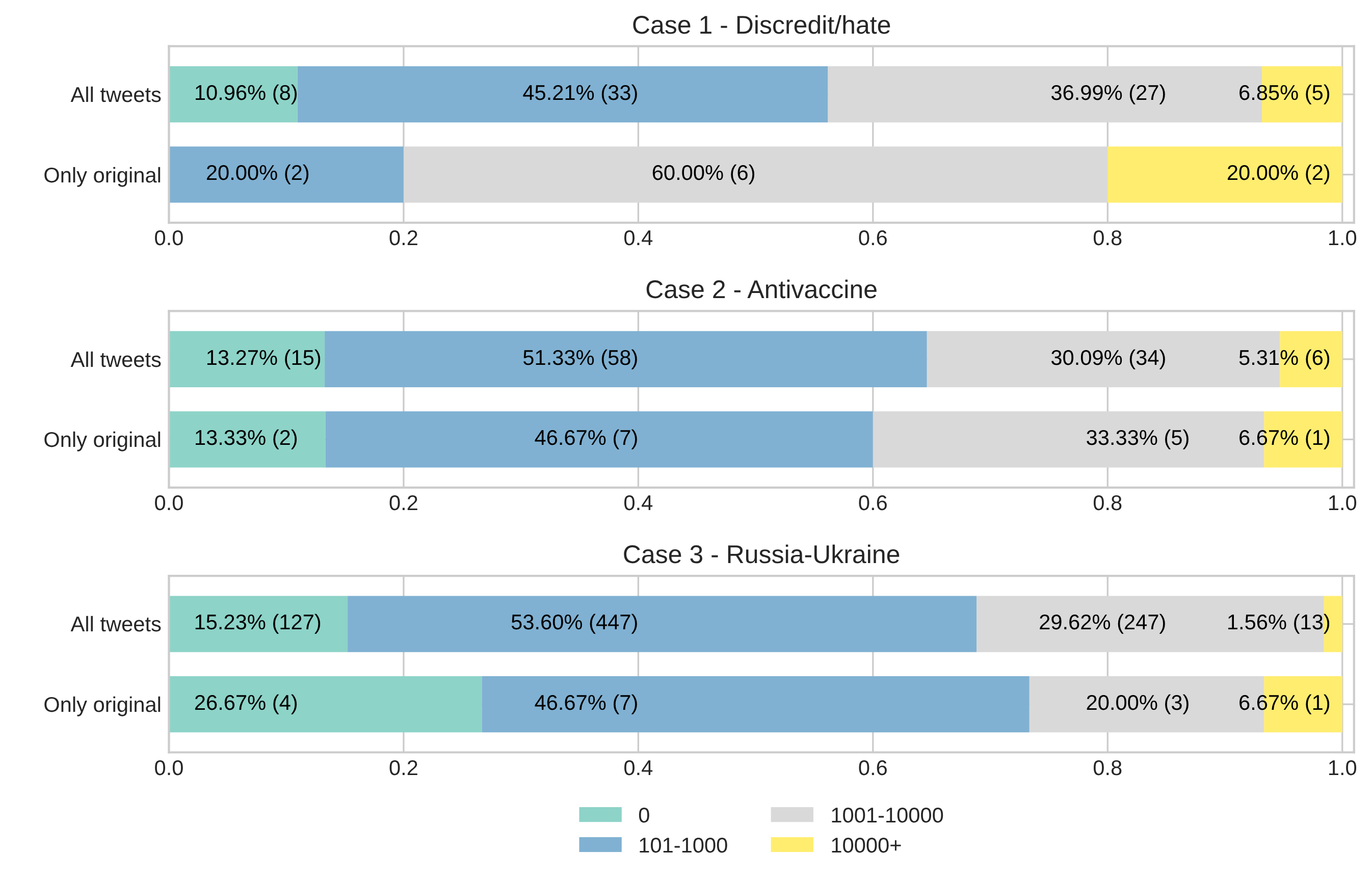}
    \captionsetup{justification=centering}
    \caption{Distribution of the groups of posts according to their number of followers, in each of the three cases.}
    \label{fig:followers}
\end{figure}

We also analyzed the number of followers of the user accounts involved in these three case studies (see Fig.~\ref{fig:followers}, showing different types of users. The original tweets in the three fluxes about misinformation are mainly shared by users from 1000 or more followers. Whereas the propagation related to the hoax about discredit/hate has more users between 1001 and 10.000 followers than those that are below or above these numbers, the other two examples indicate that authors between 101 and 1000 correspond to the largest proportion.

In the following subsections we show and describe in detail each of the three case studies. In the graphs, posts are rounded vertices, sized by the number of likes, whereas reposts are vertices with the shame of a rhombus. Posts and reposts tagged as \textit{entailment} are orange, those tagged as \textit{contradiction} are blue and \textit{neutral} ones are grey. The x-axis represents the chronological order (with annotated dates each 100 days from the first post in the visualization as a guide) and the y-axis (log scale) shows the number of followers of the authors of each post and repost.

\subsection{Case 1: Discredit/hate}
\label{case_1} 

\begin{table}[]
\centering
\resizebox{\textwidth}{!}{%
\begin{tabular}{lccccc}
\hline
\textbf{Author} & \textbf{No. Followers} & \textbf{No. interactions} & \textbf{Max No. Retweets} & \textbf{Max No. Likes} & \textbf{No. Tweets} \\ \hline
0 & 25464 & 0  & 0  & 0  & 3 \\
1 & 12537 & 7  & 7  & 5  & 4 \\
2 & 8795  & 4  & 1  & 2  & 1 \\
3 & 3881  & 1  & 1  & 0  & 1 \\
4 & 3856  & 0  & 0  & 1  & 1 \\
5 & 2669  & 0  & 0  & 1  & 1 \\
6 & 2641  & 27 & 22 & 72 & 2 \\
7 & 1141  & 1  & 1  & 2  & 1 \\
8 & 417   & 0  & 0  & 0  & 1 \\
9 & 146   & 38 & 32 & 60 & 5 \\ \hline
\end{tabular}%
}
\caption{Ranking of the most active accounts in the spread of the tweets of the case 1.}
\label{tab:users_case1}
\end{table}

\begin{figure}[!ht]
    \centering
    \includegraphics[width=1\textwidth]{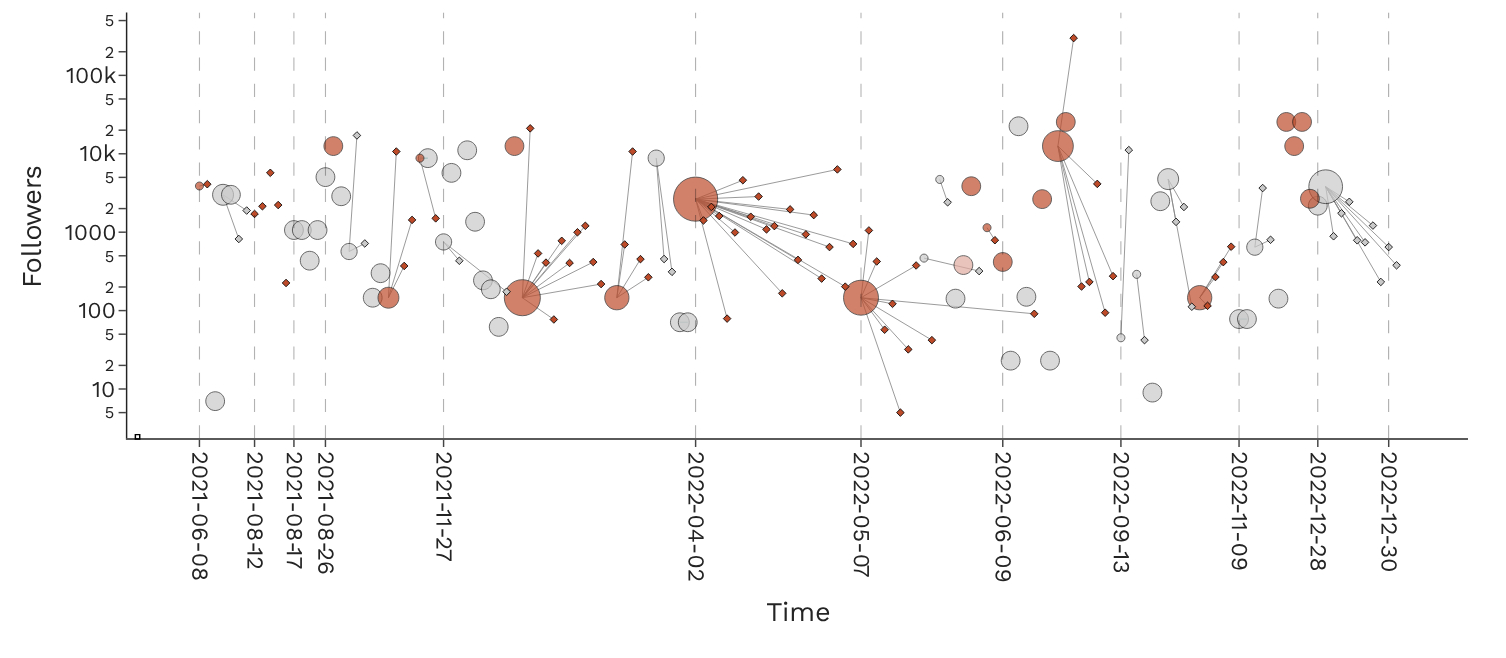}
    \captionsetup{justification=centering}
    \caption{Visualization of the graph derived from the claim ``80 percent of Muslims in Europe live from social welfare and they do not want to work''. The colour of the vertices mean blue for \textit{contradiction}, red for \textit{entailment} and gray for \textit{neutral}.}
    \label{fig:welfare}
\end{figure}

For this first case (see Fig.~\ref{fig:welfare}), we focus on a hoax that circulated on Twitter asserting that ``80 percent of Muslims living in Europe live from social welfare and they refuse to work''. For this hoax, we retrieved a large pool of information from the social network, including mentions to other users, popular or not, fake attributions to researchers to make the text trustworthy and personal views and opinions reinforcing that hoax. 

Likewise, fact-checks announce the same sentence in the negative form to combat it, as seen in the graphs generated. The variations in this case mention the name of the accounts of fact-checking and/or add links to the news debunking the hoax. It is important to note that the same statement to negate the false information is shared repeatedly at different moments in the life of the hoax.

The most retweeted post affirming the hoax took place on April 2nd, 2022, with 22 retweets, however, we can see how this false information was originally posted around 10 months before. On June 8th, 2021, an account with 3,881 followers (at the time of the data extraction) published it and received a unique retweet. It took exactly five months up to that date to see the second most retweeted post supporting this claim, with 11 retweets, and this type of misinformation resists at least until October.

The most interesting fact about the two users with more than 10.000 followers who actively wrote the contradiction of the hoax is that both of them reproduce the same action several times: one of them (25,464 followers) does it three times; the other one (12,537 followers), four. Active spreaders posting misinformation by themselves have a lower number of followers. However, in the case of retweets, there are actors with more than 10,000 followers that spread this falsehood. Maldito Bulo, from the fact-checker Maldita, emerges as the user that exceeds that number of followers as a repost of its fact-checking through another user.

This confirms that our system shows the whole cascade of misinformation and not just the static picture of the most propagated tweet in April 2022. Furthermore, this visualization also demonstrates that any virality of a post, regardless of its impact, can be preceded by posts with zero or a few active interactions (retweet or like), as shown in this case in 2021. Finally, it also reveals how after the most retweeted post supporting the claim, the flux of false information continues with different users and in different ways up to the ones shown in the last trimester of 2022.

Regarding the involvement of the specific actors that spread misinformation (see Table~\ref{tab:users_case1}), several accounts with an important number of followers posted the hoax. It is interesting to note, however, that the ones that had the most interactions, and therefore caused the most discussion around the falsehood, were some with a smaller number of followers.

\subsection{Case 2 - Antivaccine}
\label{case_2} 

The propagation of how falsely RNA vaccines against coronavirus include graphene oxide has also had an important impact on Twitter and other social media. This hoax has succeeded through paraphrasing in different periods. Whereas the previous case study showed a fixed structure and shape, this one discovers a range of posts expressing the same falsehood with different words and tones, from more declarative sentences to more aggressive ones.

\begin{table}[]
\centering
\resizebox{\textwidth}{!}{%
\begin{tabular}{lccccc}
\hline
\textbf{Author} & \textbf{No. Followers} & \textbf{No. interactions} & \textbf{Max No. Retweets} & \textbf{Max No. Likes} & \textbf{No. Tweets} \\ \hline
0 & 9547 & 26 & 23 & 23 & 1 \\
1 & 1903 & 6  & 6  & 13 & 1 \\
2 & 1854 & 8  & 1  & 4  & 3 \\
3 & 632  & 8  & 6  & 10 & 1 \\
4 & 445  & 9  & 9  & 11 & 1 \\
5 & 248  & 2  & 1  & 1  & 1 \\
6 & 158  & 11 & 9  & 11 & 1 \\
7 & 116  & 2  & 1  & 2  & 3 \\
8 & 81   & 0  & 0  & 0  & 1 \\
9 & 1    & 0  & 0  & 1  & 1 \\ \hline
\end{tabular}%
}
\caption{Ranking of the most active accounts in the spread of the tweets of the case 2.}
\label{tab:users_case2}
\end{table}

\begin{figure}[!ht]
    \centering
    \includegraphics[width=0.95\textwidth,height=0.95\textheight,keepaspectratio]{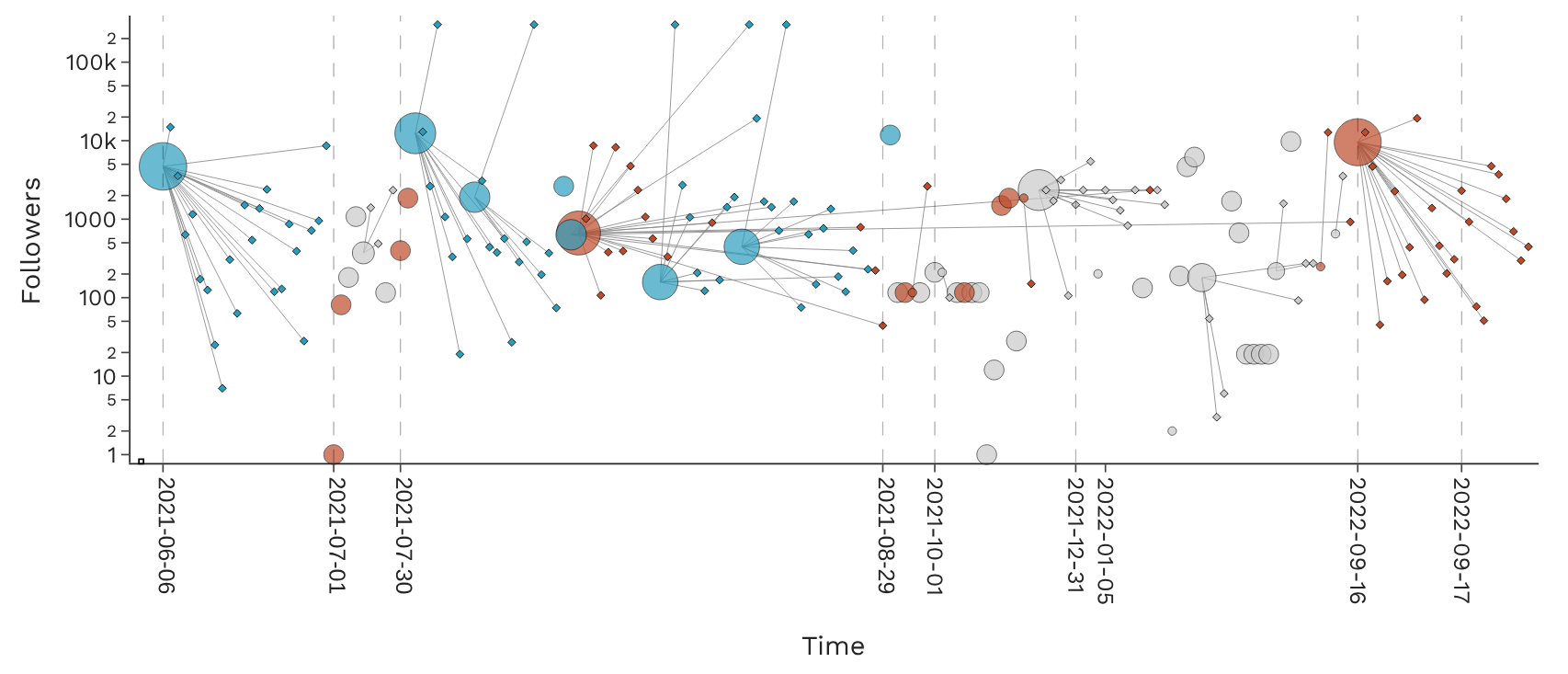}
    \captionsetup{justification=centering}
    \caption{Visualization of the graph derived from the claim ``Messenger RNA vaccines against COVID-19 contain graphene oxide''.}
    \label{fig:graphene}
\end{figure}

Again, fact-checks are mostly composed of the same sentence, including small variations mentioning the fact-checking source and/or the link. In some specific cases, mainly in the first interactions of fact-checkers to contradict false information, we can observe longer explanations, substantiating the falsity of the fact. 

In this particular case study, we can see that the first tweets composing the propagation cascade deny the hoax. This surprising effect is present in those cascades where the origin is in other environments (i.e., chain messages on WhatsApp). The chain of propagation originates on 6 June 2021, and about a month later we can observe some tweets disseminating the hoax. It is noteworthy that the tweet that has the greatest effect on the dissemination of the hoax (16 September 2022) occurs after months in which the only references to the hoax are \textit{neutral}.


Analyzing in more detail the accounts involved (see Table~\ref{case_2}), the two accounts that typed the contradiction of the claim have more than 10,000 followers, but none with this influence actively expresses the hoax itself (the most influential one has 9,547 followers). Close to these numbers, the last original tweet sharing this piece of misinformation had 9,547 followers. Regarding retweets, Maldito Bulo, from the fact-checker Maldita (300,081 followers at that moment) played an important role in this hoax.

This case study reveals the limitations of traditional methods, which might have analyzed this publication in August or its final iteration in September, without considering the origins or evolution of the misinformation involved. By adopting this new approach, we gain insight into the broader context of how misinformation spreads. It shows that the groups opposing a piece of false information represent just a segment of the overall dissemination of posts linked to misinformation, as it continues to spread among other individuals who either refute or challenge the assertion. The spread of the latest hoax on September 16, which occurred without any users debunking it, underscores this point. This example highlights the complexity of misinformation propagation and the necessity of considering its full lifecycle for effective analysis.

\subsection{Case 3: Russia-Ukraine}
\label{case_3}

\begin{figure}[!ht]
    \centering
    \includegraphics[width=0.95\textwidth,height=0.95\textheight,keepaspectratio]{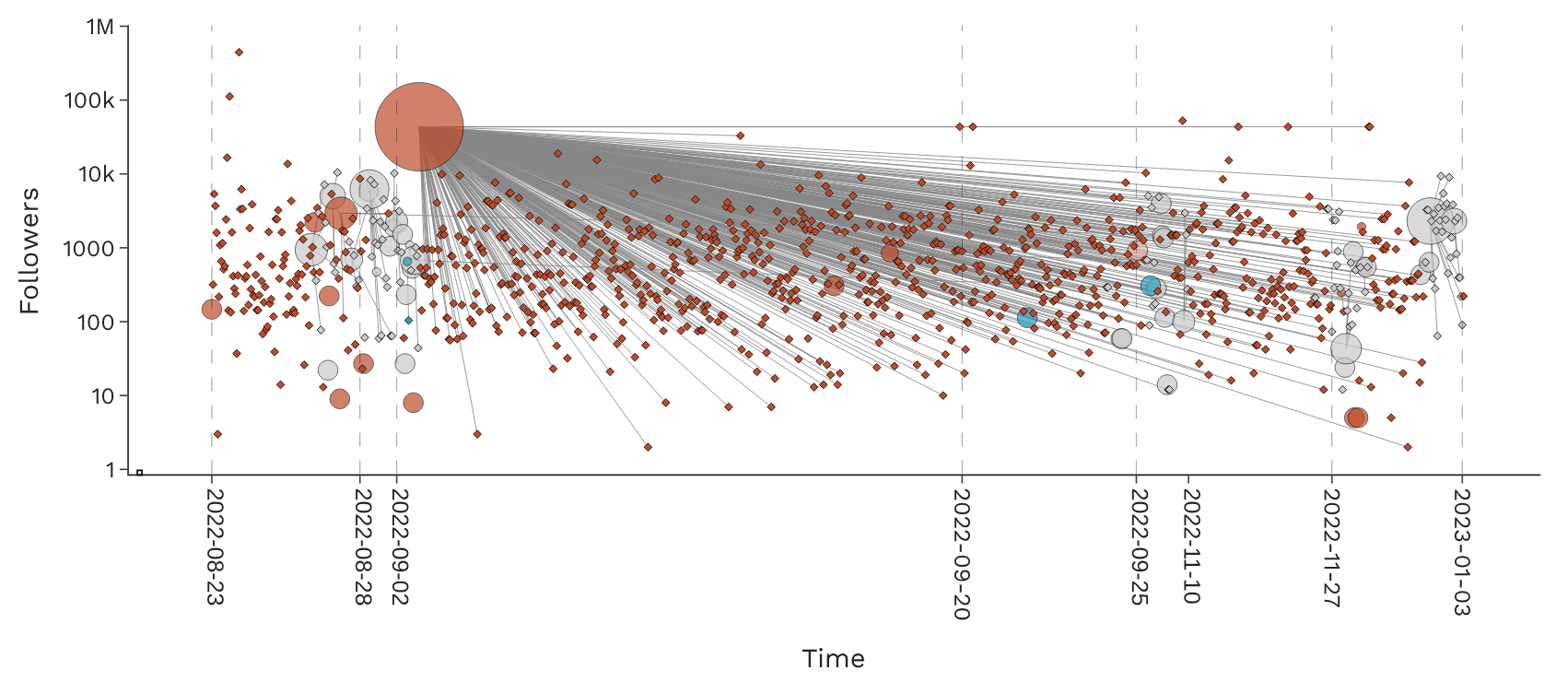}
    \captionsetup{justification=centering}
    \caption{Visualization of the graph derived from the claim ``Zelensky sold 17 million hectares of land to Monsanto, Dupont and Cargill''.}
    \label{fig:zelensky}
\end{figure}

This third case, shown in Fig.~\ref{fig:zelensky}, considers the hoax ``Zelensky sold 17 million hectares of land to Monsanto, Dupont and Cargill''. Unlike the earlier cases, the visualizations here highlight variations in impact, illustrating the hoax's dissemination through a multitude of messages. Echoing the patterns seen in the previous example, there is not a single form of post; instead, a diversity of presentations emerge, ranging from the use of hashtags and user mentions to varying styles and tactics to engage the audience. The spread of this misinformation is primarily driven by reposts, yet these varied expressions of the same false claim also play a significant role, particularly at the initial stages of its spread. This multiplicity of formats and channels underscores the complex nature of misinformation propagation and the challenges in tracing and countering it.

This example showcases a viral post from September 19th, 2022, that stood out with 663 retweets and 1,000 likes. Remarkably, its retweets occurred not just on the day of the original tweet but continued sporadically until December 30th of the same year, although diminishing impact. This trend is visualized along the x-axis and highlights both the immediate impact of the hoax and its prolonged presence in the digital discourse.

Thanks to this visualization, abnormal activity is shown on the same user who writes the viral post. This suggests that this person retweets it several times, according to the download of Twitter API data. This continuous activity preserves the propagation of this false information and prevents it from dying on Twitter (X). In addition, there was another post with even more retweets whose initial user has been deleted or suspended (1.597 retweets), as there are no edges linking them to the original post.

\begin{table}[]
\centering
\resizebox{\textwidth}{!}{%
\begin{tabular}{cccccc}
\hline
\textbf{Author} & \textbf{No. Followers} & \textbf{No. Interactions} & \textbf{Max No. Retweets} & \textbf{Max No. Likes} & \textbf{No. Tuits} \\ \hline
0 & 43502 & 774 & 663 & 1000 & 1 \\
1 & 2933  & 8   & 7   & 11   & 1 \\
2 & 2211  & 0   & 0   & 1    & 1 \\
3 & 1925  & 3   & 1   & 1    & 1 \\
4 & 843   & 2   & 2   & 4    & 1 \\
5 & 248   & 2   & 1   & 4    & 1 \\
6 & 223   & 0   & 0   & 1    & 1 \\
7 & 147   & 0   & 0   & 0    & 1 \\
8 & 113   & 1   & 0   & 1    & 1 \\
9 & 8     & 0   & 0   & 0    & 1 \\ \hline
\end{tabular}%
}
\caption{Ranking of the most active accounts in the spread of the tweets of the case 3.}
\label{tab:users_case3}
\end{table}

In this case, the virality of the post on September 19th already shows the vast propagation of this misinformation, how it arrives to other users throughout time, how it can be the continuation of other viral posts in the past, and even the strategies led by users to keep the impact of a tweet. This viral post has the same publication date as others with their own content that also spread content about the hoax. In Table~\ref{tab:users_case3}, it can be seen how one user is the one leading the whole propagation cascade, spreading the hoax while being the center of a high number of interactions.


\section{Results and discussion}
\label{results_and_discussion}

In every analyzed example, DisTrack answers positively to the question ``Can we extract the conversation about a hoax on Twitter?''. The conversations about the three chosen falsehoods have been visualized after downloading and processing and refining the data and metadata of the posts related to them. This demonstrates the success in generating automated queries to extract as many tweets as possible about false claims to later filter them depending on their type of alignment (\textit{entailment}, \textit{contradiction} or \textit{neutral}).

This leads to the affirmative answer to the second subquestion ``Can we separate tweets related to the hoax in the extracted conversation from tweets not related to it?'') and to the third one (``Can we distinguish between hoaxes that propagate a hoax from those that contradict it?''). The application of NLI leveraged with Transformers has been satisfactory at two levels: not only by separating tweets that are not related to each false claim, but also by separating the \textit{entailment} posts (those that state misinformation) from \textit{contradiction} posts (those that deny them).

In particular, the use of Transformers in the application of NLI to detect misinformation in this case has allowed DisTrack to identify falsehoods on X, regardless of the way they were exposed in the platform. In case 1, the posts were similar to the false claim taken as the reference, with subtle changes, but in cases 2 and 3, the posts paraphrased the content of the false claim together with hashtags, links or other elements, and their \textit{entailment} was also guessed by the NLI model. In these downloaded conversations about misinformation, The existence of accounts reproducing exactly the same original content and of those that paraphrase it instead encourages research to unveil the dynamics of falsehoods and the existence of bots in contexts of crises~\cite{himelein2021bots}. 

Thanks to the whole process of searching tweets, aligning them with the validated claim and visual representation, what we present is a complete architecture that allows the automated and detailed analysis of the spread of disinformation on a social network. From the first references to the hoax, to the nodes that most influence its dissemination (thanks to their number of followers, for example) or the patterns observed in the cascade, DisTrack allows us to understand first-hand the dynamics of disinformation and how it spreads and generates a certain impact.

The fourth research question ``Can we identify the users involved in the conversation of a hoax from beginning to end?'' is also answered affirmatively. DisTrack shows a closer image of the whole life of a piece of misinformation: viral tweets matter, but all the participants in the conversation are relevant. In the three case studies analyzed, falsehoods appear before and after the spread of the most retweeted posts, in the shape of less viral tweets. This research reveals the characteristics of every actor inside the ecosystem of false information: the case studies unveil users with many followers as actors in the propagation of a hoax, those conceived as 'super-spreaders'~\cite{bodaghi2022theater, carrasco2021participacion}, but also accounts with different weights in terms of followers, not always the most followed ones~\cite{himelein2021bots}.

The results of the four subquestions allow us to answer the main research question: ``Is it possible to track conversations around specific hoaxes on Twitter (X)?''. The three use cases confirm that misinformation can be traced on Twitter by modeling it, using the three presented modules: firstly, the tweets of the conversation about falsehoods were extracted through automated queries (subquestion 1); secondly, they were separated from each other according to their alignment to that false information (subquestions 2 and 3), and, finally, their authors and their characteristics were also identified (subquestion 4) through the final generation of graphs that show the evolution of that falsehood.

These outcomes follow recent research relying on tweet extraction and NLI-based classification in the field of misinformation. These studies show the orchestration of posts with different ranges of influence contributing to a more complex propagation, and also the different nature of users in the conversation about specific falsehoods, in accordance with the results provided by DisTrack. The exploratory data analysis of the three presented use cases was a demonstration of this variety of posts and users.

Nevertheless, DisTrack adds a layer of research through the generation of the graphs to exploit more the synergies between NLP and SNA as the fields in charge of combating misinformation~\cite{montoro2023fighting}: a falsehood does not die progressively after its more viral publication and is not always born directly from it, and fact-checks are also repeated in the successive lives of each falsehood. In the cases analyzed, the conversation is reshaped and hoaxes arise again. Furthermore, the evolution of tweets from fact-checkers and of those that refuse verbally the selected hoaxes does not only occur with the virality of that type of misinformation at its best, but also in other periods, as observed in the final visualizations of this research. 

The combination of falsehoods and the posts that debunk them in a final visualization is also worth mentioning. DisTrack proposes an additional step to the precursory work: whereas previous experiments contribute to a better knowledge of the nature of the posts about misinformation, they do not offer any surveillance to monitor and mitigate it. Nevertheless, DisTrack brings back those steps of the tweets extraction and NLI filtering to put them at the service of graph generation for that desired supervision. This enables the coordinated response of fact-checkers and the control of their effects on false information.

Overall, DisTrack reinforces the need to study misinformation as a viral model~\cite{goel2016structural} in a chain of infections by posts with different ranges of influence. Its opposed model covered in research, the broadcast model, would only have shown a part of misinformation because it only conceives contagion as a unique primary vertex infecting the rest. On the contrary, the examples made by Distrack print several versions of the vertex, various infections and, as a consequence, many propagations instead of one. 

This is also important in the combination of \textit{entailment} and \textit{contradiction} posts. Not only do broadcast models~\cite{goel2016structural} isolate a cascade of misinformation from the rest of the generated false information, but they also put it apart from the posts that contradict it and from fact-checkers, preventing researchers from depicting their appearances in the ecosystem of that cascade. Likewise, a broadcast-model cascade of fact-checks without showing the rest of the conversation through the steps of DisTrack would remove them from the falsehoods they counteract.

\section{Conclusions}
\label{conclusions}

All in all, our work releases a line of action through the shape of DisTrack, as the beginning of a tool able to merge Transformer-leveraged tweet extraction, NLI-driven tagging of misinformation from the posts retrieved and graph generation of Twitter-based and user-based properties in an output that shows chronologically the evolution of a conversation motivated by misinformation (spreaders, fact-checkers and other users) across the different publications and actors involved. 

With this proposed line of action, future experiments can study how DisTrack modules can be modified. With NLP as one of the core parts of this research, there is room for improvement given the advance of new Language Models (LM): on the one hand, with their application for new models that capture better the topics and, thus, the keywords for the query that enable the download of posts; on the other hand, with their use for NLI to increase the level of accuracy in the classification of posts as \textit{entailment}, \textit{contradiction} of \textit{neutral}, the three tags used to colour the vertices in the final graphs to build the representative picture of misinformation.

Future work includes the application of DisTrack beyond misinformation. For instance, previous research shows how sentiment analysis has become relevant in the studies about aggressive discourse in the context of government elections ~\cite{torregrosa2023mixed} and has stated Twitter as a ``sentiment thermometer'' through VADER ~\cite{hutto2014vader}. Although the first steps of these experiments evoque the implementation of DisTrack, involving the download of posts from X and the extraction of features for the study of specific behaviours, the part of monitoring polarization through the combination of NLP and SNA through graph generation is missing, unlike DisTrack. For this reason, future experiments could be oriented to develop the same modules as DisTrack but by substituting the task of detecting false information in favour of analyzing sentiment in this platform, a space with more proportion of politicians and political activists than others~\cite{newman2023digital}.

Accordingly, this research opens the door to the development of the part of tracking in other scenarios, where NLI does not have to be necessarily excluded and can enrich the information in the graph (in the previous case, for example, to know the users stating the same political information and, thus, enunciating the same message to the audience for the elections). These other scenarios can include: a more general approach, beyond politics, to take more advantage of the advances of Transformers for hate speech detection ~\cite{mutanga2020hate, roy2021leveraging} and follow the trajectory of harmful posts on social media; an alternative approach in the domain of author profiling, where graphs can leverage recent research in LLM for this task~\cite{huertas2024understanding} by tracing information and linking the vertices that are likely to have the same attribution, or the area of topic modeling, fueled by models such as BERTopic~\cite{grootendorst2022bertopic}, in which the concatenation of the modules proposed by DisTrack could result in edges tying the posts about the same issue, among other fields of study.

DisTrack also serves as an initiative to explore platforms beyond Twitter (X). The international survey developed by Reuters Institute Digital News Report in 2022 already revealed the prevalence of Facebook, YouTube, Instagram or TikTok as well as WhatsApp, Telegram or Facebook Messenger as ways of consuming news~\cite{holig2022reuters}, which has continued in 2023~\cite{newman2023digital}. Each of them has a different structure but DisTrack arises as a proposal to be adapted to other scenarios of misinformation on these ecosystems.

Although OSNs mutate or change, with Twitter, now X, as an example of that, users continue searching platforms to be in the circles of their ecosystems. Research about migrations from Twitter to Bluesky, Mastodon and Threads has already been covered~\cite{jeong2024user}, showing the interest in the current microblogging social media and, thus, the necessity of fighting against their information disorders, motivated by tools and methodologies such as DisTrack.


\section*{Declarations}

\subsection*{\textbf{Competing Interests}}

The authors declare no competing interests.

\subsection*{\textbf{Funding}}

This work has been funded by the project PCI2022-134990-2 (MARTINI) of the CHISTERA IV Cofund 2021 program, funded by MCIN/AEI/10.13039/ 501100011033 and by the ``European Union NextGenerationEU/PRTR''; by the research project DisTrack: Tracking disinformation in Online Social Networks through Deep Natural Language Processing, granted by Mobile World Capital Foundation; by the Spanish Ministry of Science and Innovation under FightDIS (PID2020-117263GB-I00); by MCIN/AEI/10.13039/501100011033/ and European Union NextGenerationEU/PRTR for XAI-Disinfodemics (PLEC 2021-007681) grant, by European Comission under IBERIFIER Plus - Iberian Digital Media Observatory (DIGITAL-2023-DEPLOY- 04-EDMO-HUBS 101158511); and by EMIF managed by the Calouste Gulbenkian Foundation, in the project MuseAI.

\bibliographystyle{elsarticle-num}

\bibliography{bibliography.bib}

\begin{thebibliography}{10}
\expandafter\ifx\csname url\endcsname\relax
  \def\url#1{\texttt{#1}}\fi
\expandafter\ifx\csname urlprefix\endcsname\relax\def\urlprefix{URL }\fi
\expandafter\ifx\csname href\endcsname\relax
  \def\href#1#2{#2} \def\path#1{#1}\fi

\bibitem{salaverria2020desinformacion}
R.~Salaverr{\'\i}a, N.~Busl{\'o}n, F.~L{\'o}pez-Pan, B.~Le{\'o}n, I.~L{\'o}pez-Go{\~n}i, M.-C. Erviti, Desinformaci{\'o}n en tiempos de pandemia: tipolog{\'\i}a de los bulos sobre la covid-19, Profesional de la Informaci{\'o}n 29~(3) (2020).

\bibitem{said2021evolucion}
E.~M. Said-Hung, M.~A. Merino-Arribas, J.~Mart{\'\i}nez-Torres, Evoluci{\'o}n del debate acad{\'e}mico en la web of science y scopus sobre unfaking news (2014-2019), Estudios Sobre el Mensaje Period{\'\i}stico 27~(3) (2021) 961.

\bibitem{guess2020misinformation}
A.~M. Guess, B.~A. Lyons, Misinformation, disinformation, and online propaganda, Social media and democracy: The state of the field, prospects for reform 10 (2020).

\bibitem{karlova2013social}
N.~A. Karlova, K.~E. Fisher, A social diffusion model of misinformation and disinformation for understanding human information behaviour, Information Research (2013).

\bibitem{wardle2017information}
C.~Wardle, H.~Derakhshan, Information disorder: Toward an interdisciplinary framework for research and policymaking, Vol.~27, Council of Europe Strasbourg, 2017.

\bibitem{ireton2018journalism}
C.~Ireton, J.~Posetti, Journalism, fake news \& disinformation: handbook for journalism education and training, Unesco Publishing, 2018.

\bibitem{posetti2018short}
J.~Posetti, A.~Matthews, A short guide to the history of ‘fake news’ and disinformation, International Center for Journalists 7~(2018) (2018) 2018--07.

\bibitem{kouzy2020coronavirus}
R.~Kouzy, J.~Abi~Jaoude, A.~Kraitem, M.~B. El~Alam, B.~Karam, E.~Adib, J.~Zarka, C.~Traboulsi, E.~W. Akl, K.~Baddour, Coronavirus goes viral: quantifying the covid-19 misinformation epidemic on twitter, Cureus 12~(3) (2020).

\bibitem{newman2023digital}
N.~Newman, R.~Fletcher, K.~Eddy, C.~T. Robertson, R.~K. Nielsen, Digital news report 2023, RISJ: Reuters Institute for the Study of Journalism (2023).

\bibitem{choras2021advanced}
M.~Chora{\'s}, K.~Demestichas, A.~Gie{\l}czyk, {\'A}.~Herrero, P.~Ksieniewicz, K.~Remoundou, D.~Urda, M.~Wo{\'z}niak, Advanced machine learning techniques for fake news (online disinformation) detection: A systematic mapping study, Applied Soft Computing 101 (2021) 107050.

\bibitem{freelon2020disinformation}
D.~Freelon, C.~Wells, Disinformation as political communication, Political communication 37~(2) (2020) 145--156.

\bibitem{altaysurvey}
S.~Altay, M.~Berriche, H.~Heuer, J.~Farkas, S.~Rathje, A survey of expert views on misinformation: Definitions, determinants, solutions, and future of the field, Harvard Kennedy School Misinformation Review 4~(4) (2023) 1--34.

\bibitem{brennen2020types}
J.~S. Brennen, F.~M. Simon, P.~N. Howard, R.~K. Nielsen, Types, sources, and claims of covid-19 misinformation, Reuters Institute for the Study of Journalism (2020).

\bibitem{martin2022facter}
A.~Mart{\'\i}n, J.~Huertas-Tato, {\'A}.~Huertas-Garc{\'\i}a, G.~Villar-Rodr{\'\i}guez, D.~Camacho, Facter-check: Semi-automated fact-checking through semantic similarity and natural language inference, Knowledge-Based Systems (2022) 109265.

\bibitem{hasan2021review}
I.~Hasan, S.~Rizvi, Review of ai techniques and cognitive computing framework for intelligent decision support, in: 2021 8th International Conference on Computing for Sustainable Global Development (INDIACom), IEEE, 2021, pp. 891--898.

\bibitem{devlin2018bert}
J.~Devlin, M.-W. Chang, K.~Lee, K.~Toutanova, Bert: Pre-training of deep bidirectional transformers for language understanding, arXiv preprint arXiv:1810.04805 (2018).

\bibitem{liu2019roberta}
Y.~Liu, M.~Ott, N.~Goyal, J.~Du, M.~Joshi, D.~Chen, O.~Levy, M.~Lewis, L.~Zettlemoyer, V.~Stoyanov, Roberta: A robustly optimized bert pretraining approach, arXiv preprint arXiv:1907.11692 (2019).

\bibitem{lample2019cross}
G.~Lample, A.~Conneau, Cross-lingual language model pretraining, arXiv preprint arXiv:1901.07291 (2019).

\bibitem{mikolov2013distributed}
T.~Mikolov, I.~Sutskever, K.~Chen, G.~S. Corrado, J.~Dean, Distributed representations of words and phrases and their compositionality, Advances in neural information processing systems 26 (2013).

\bibitem{pennington2014glove}
J.~Pennington, R.~Socher, C.~D. Manning, Glove: Global vectors for word representation, in: Proceedings of the 2014 conference on empirical methods in natural language processing (EMNLP), 2014, pp. 1532--1543.

\bibitem{tretiakov2022detection}
A.~Tretiakov, A.~Mart{\'\i}n, D.~Camacho, Detection of false information in spanish using machine learning techniques, in: International Conference on Intelligent Data Engineering and Automated Learning, Springer, 2022, pp. 42--53.

\bibitem{jwa2019exbake}
H.~Jwa, D.~Oh, K.~Park, J.~M. Kang, H.~Lim, exbake: Automatic fake news detection model based on bidirectional encoder representations from transformers (bert), Applied Sciences 9~(19) (2019) 4062.

\bibitem{montoro2023fighting}
A.~Montoro-Montarroso, J.~Cant{\'o}n-Correa, P.~Rosso, B.~Chulvi, {\'A}.~Panizo-Lledot, J.~Huertas-Tato, B.~Calvo-Figueras, M.~J. Rementeria, J.~G{\'o}mez-Romero, Fighting disinformation with artificial intelligence: fundamentals, advances and challenges, Profesional de la informaci{\'o}n 32~(3) (2023).

\bibitem{vijjali2020two}
R.~Vijjali, P.~Potluri, S.~Kumar, S.~Teki, Two stage transformer model for covid-19 fake news detection and fact checking, arXiv preprint arXiv:2011.13253 (2020).

\bibitem{huertas2021civic}
{\'A}.~Huertas-Garc{\'\i}a, J.~Huertas-Tato, A.~Mart{\'\i}n, D.~Camacho, Civic-upm at checkthat! 2021: Integration of transformers in misinformation detection and topic classification., in: CLEF (Working Notes), 2021, pp. 520--530.

\bibitem{huertas2021countering}
{\'A}.~Huertas-Garc{\'\i}a, J.~Huertas-Tato, A.~Mart{\'\i}n, D.~Camacho, Countering misinformation through semantic-aware multilingual models, in: International conference on intelligent data engineering and automated learning, Springer, 2021, pp. 312--323.

\bibitem{gaglani2020unsupervised}
J.~Gaglani, Y.~Gandhi, S.~Gogate, A.~Halbe, Unsupervised whatsapp fake news detection using semantic search, in: 2020 4th International Conference on Intelligent Computing and Control Systems (ICICCS), IEEE, 2020, pp. 285--289.

\bibitem{guo2020cord19sts}
X.~Guo, H.~Mirzaalian, E.~Sabir, A.~Jaiswal, W.~Abd-Almageed, Cord19sts: Covid-19 semantic textual similarity dataset, arXiv preprint arXiv:2007.02461 (2020).

\bibitem{larraz2023semantic}
I.~Larraz, F.~Sallicati, et~al., Semantic similarity models for automated fact-checking: Claimcheck as a claim matching tool, Profesional de la informaci{\'o}n 32~(3) (2023).

\bibitem{maccartney2009natural}
B.~MacCartney, Natural language inference, Stanford University, 2009.

\bibitem{gururangan2018annotation}
S.~Gururangan, S.~Swayamdipta, O.~Levy, R.~Schwartz, S.~R. Bowman, N.~A. Smith, Annotation artifacts in natural language inference data, arXiv preprint arXiv:1803.02324 (2018).

\bibitem{bowman2015large}
S.~R. Bowman, G.~Angeli, C.~Potts, C.~D. Manning, A large annotated corpus for learning natural language inference, arXiv preprint arXiv:1508.05326 (2015).

\bibitem{williams2017broad}
A.~Williams, N.~Nangia, S.~R. Bowman, A broad-coverage challenge corpus for sentence understanding through inference, arXiv preprint arXiv:1704.05426 (2017).

\bibitem{conneau2018xnli}
A.~Conneau, G.~Lample, R.~Rinott, A.~Williams, S.~R. Bowman, H.~Schwenk, V.~Stoyanov, Xnli: Evaluating cross-lingual sentence representations, arXiv preprint arXiv:1809.05053 (2018).

\bibitem{huertas2021sml}
J.~Huertas-Tato, A.~Mart{\'\i}n, D.~Camacho, Silt: Efficient transformer training for inter-lingual inference, Expert Systems with Applications 200 (2022) 116923.

\bibitem{camacho2020four}
D.~Camacho, {\'A}.~Panizo-LLedot, G.~Bello-Orgaz, A.~Gonzalez-Pardo, E.~Cambria, The four dimensions of social network analysis: An overview of research methods, applications, and software tools, Information Fusion 63 (2020) 88--120.

\bibitem{panizo2019describing}
A.~Panizo-LLedot, J.~Torregrosa, G.~Bello-Orgaz, J.~Thorburn, D.~Camacho, Describing alt-right communities and their discourse on twitter during the 2018 us mid-term elections, in: International conference on complex networks and their applications, Springer, 2019, pp. 427--439.

\bibitem{tacchini2017some}
E.~Tacchini, G.~Ballarin, M.~L. Della~Vedova, S.~Moret, L.~De~Alfaro, Some like it hoax: Automated fake news detection in social networks, arXiv preprint arXiv:1704.07506 (2017).

\bibitem{sharma2019combating}
K.~Sharma, F.~Qian, H.~Jiang, N.~Ruchansky, M.~Zhang, Y.~Liu, Combating fake news: A survey on identification and mitigation techniques, ACM Transactions on Intelligent Systems and Technology (TIST) 10~(3) (2019) 1--42.

\bibitem{parikh2018media}
S.~B. Parikh, P.~K. Atrey, Media-rich fake news detection: A survey, in: 2018 IEEE conference on multimedia information processing and retrieval (MIPR), IEEE, 2018, pp. 436--441.

\bibitem{vosoughi2018spread}
S.~Vosoughi, D.~Roy, S.~Aral, The spread of true and false news online, science 359~(6380) (2018) 1146--1151.

\bibitem{saby2021twitter}
D.~Saby, O.~Philippe, N.~Busl{\'o}n, J.~del Valle, O.~Puig, R.~Salaverr{\'\i}a, M.~J. Rementeria, Twitter analysis of covid-19 misinformation in spain, in: Computational Data and Social Networks: 10th International Conference, CSoNet 2021, Virtual Event, November 15--17, 2021, Proceedings 10, Springer, 2021, pp. 267--278.

\bibitem{bello2017detecting}
G.~Bello-Orgaz, J.~Hernandez-Castro, D.~Camacho, Detecting discussion communities on vaccination in twitter, Future Generation Computer Systems 66 (2017) 125--136.

\bibitem{goel2016structural}
S.~Goel, A.~Anderson, J.~Hofman, D.~J. Watts, The structural virality of online diffusion, Management Science 62~(1) (2016) 180--196.

\bibitem{bodaghi2022theater}
A.~Bodaghi, J.~Oliveira, The theater of fake news spreading, who plays which role? a study on real graphs of spreading on twitter, Expert Systems with Applications 189 (2022) 116110.

\bibitem{carrasco2021participacion}
R.~Carrasco~Polaino, M.~{\'A}. Mart{\'\i}n~C{\'a}rdaba, E.~Villar~Cirujano, Participaci{\'o}n ciudadana en twitter. pol{\'e}micas anti-vacunas en tiempos de covid-19, Comunicar: Revista cient{\'\i}fica iberoamericana de comunicaci{\'o}n y educaci{\'o}n.(Ejemplar dedicado a: Participaci{\'o}n ciudadana en la esfera digital) 29~(69) (2021) 21--31.

\bibitem{villar2022virality}
G.~Villar-Rodr{\'\i}guez, M.~Souto-Rico, A.~Mart{\'\i}n, Virality, only the tip of the iceberg: ways of spread and interaction around covid-19 misinformation in twitter, Communication \& Society (2022) 239--256.

\bibitem{noguera2023disinformation}
J.~M. Noguera-Vivo, M.~del Mar Grand{\'\i}o-P{\'e}rez, G.~Villar-Rodr{\'\i}guez, A.~Mart{\'\i}n, D.~Camacho, Disinformation and vaccines on social networks: Behavior of hoaxes on twitter, Revista Latina de Comunicaci{\'o}n Social~(81) (2023) 44--62.

\bibitem{grootendorst2020keybert}
M.~Grootendorst, \href{https://doi.org/10.5281/zenodo.4461265}{Keybert: Minimal keyword extraction with bert.} (2020).
\newblock \href {https://doi.org/10.5281/zenodo.4461265} {\path{doi:10.5281/zenodo.4461265}}.
\newline\urlprefix\url{https://doi.org/10.5281/zenodo.4461265}

\bibitem{conneau2019unsupervised}
A.~Conneau, K.~Khandelwal, N.~Goyal, V.~Chaudhary, G.~Wenzek, F.~Guzm{\'a}n, E.~Grave, M.~Ott, L.~Zettlemoyer, V.~Stoyanov, Unsupervised cross-lingual representation learning at scale, arXiv preprint arXiv:1911.02116 (2019).

\bibitem{nie2019adversarial}
Y.~Nie, A.~Williams, E.~Dinan, M.~Bansal, J.~Weston, D.~Kiela, Adversarial nli: A new benchmark for natural language understanding, arXiv preprint arXiv:1910.14599 (2019).

\bibitem{thorne2018fever}
J.~Thorne, A.~Vlachos, C.~Christodoulopoulos, A.~Mittal, Fever: a large-scale dataset for fact extraction and verification, arXiv preprint arXiv:1803.05355 (2018).

\bibitem{kinga2015method}
D.~Kinga, J.~B. Adam, et~al., A method for stochastic optimization, in: International conference on learning representations (ICLR), Vol.~5, San Diego, California;, 2015, p.~6.

\bibitem{himelein2021bots}
M.~Himelein-Wachowiak, S.~Giorgi, A.~Devoto, M.~Rahman, L.~Ungar, H.~A. Schwartz, D.~H. Epstein, L.~Leggio, B.~Curtis, Bots and misinformation spread on social media: Implications for covid-19, Journal of medical Internet research 23~(5) (2021) e26933.

\bibitem{torregrosa2023mixed}
J.~Torregrosa, S.~D’Antonio-Maceiras, G.~Villar-Rodr{\'\i}guez, A.~Hussain, E.~Cambria, D.~Camacho, A mixed approach for aggressive political discourse analysis on twitter, Cognitive computation 15~(2) (2023) 440--465.

\bibitem{hutto2014vader}
C.~Hutto, E.~Gilbert, Vader: A parsimonious rule-based model for sentiment analysis of social media text, in: Proceedings of the international AAAI conference on web and social media, Vol.~8, 2014, pp. 216--225.

\bibitem{mutanga2020hate}
R.~T. Mutanga, N.~Naicker, O.~O. Olugbara, Hate speech detection in twitter using transformer methods, International Journal of Advanced Computer Science and Applications 11~(9) (2020).

\bibitem{roy2021leveraging}
S.~G. Roy, U.~Narayan, T.~Raha, Z.~Abid, V.~Varma, Leveraging multilingual transformers for hate speech detection, arXiv preprint arXiv:2101.03207 (2021).

\bibitem{huertas2024understanding}
J.~Huertas-Tato, A.~Mart{\'\i}n, D.~Camacho, Understanding writing style in social media with a supervised contrastively pre-trained transformer, Knowledge-Based Systems 296 (2024) 111867.

\bibitem{grootendorst2022bertopic}
M.~Grootendorst, Bertopic: Neural topic modeling with a class-based tf-idf procedure, arXiv preprint arXiv:2203.05794 (2022).

\bibitem{holig2022reuters}
S.~H{\"o}lig, J.~Behre, W.~Schulz, Reuters institute digital news report 2022: Ergebnisse f{\"u}r deutschland, Reuters Institute (2022).

\bibitem{jeong2024user}
U.~Jeong, A.~Nirmal, K.~Jha, S.~X. Tang, H.~R. Bernard, H.~Liu, User migration across multiple social media platforms, in: Proceedings of the 2024 SIAM International Conference on Data Mining (SDM), SIAM, 2024, pp. 436--444.

\end{thebibliography}

\end{document}